\documentclass[journal,draftclsnofoot,onecolumn,12pt,letterpaper]{IEEEtran}

\usepackage{graphicx}
\usepackage[cmex10]{amsmath}
\usepackage{mathtools}
\usepackage{cite}
\ifCLASSOPTIONcompsoc
\usepackage[caption=false,font=normalsize,labelfont=sf,textfont=sf]{subfig}
\else
\usepackage[caption=false,font=footnotesize]{subfig}
\fi

\begin{document}
%
\title{Stability Region of a Slotted Aloha Network \\ with K-Exponential Backoff}
%
%

\author{Farzaneh~Farhadi,~\IEEEmembership{Student~Member,~IEEE,}
        and~Farid~Ashtiani,~\IEEEmembership{Member,~IEEE}}
        

\maketitle

\ifCLASSOPTIONdraftclsnofoot
\vspace{-1.2cm}
\fi
\begin{abstract}
Stability region of random access wireless networks is known for only simple network scenarios. The main problem in this respect is due to interaction among queues. When transmission probabilities during successive transmissions change, e.g., when exponential backoff mechanism is exploited, the interactions in the network are stimulated. In this paper, we derive the stability region of a buffered slotted Aloha network with K-exponential backoff mechanism, approximately, when a finite number of nodes exist. To this end, we propose a new approach in modeling the interaction among wireless nodes. In this approach, we model the network with inter-related quasi-birth-death (QBD) processes such that at each QBD corresponding to each node, a finite number of phases consider the status of the other nodes. Then, by exploiting the available theorems on stability of QBDs, we find the stability region. We show that exponential backoff mechanism is able to increase the area of the stability region of a simple slotted Aloha network with two nodes, more than 40\%. We also show that a slotted Aloha network with exponential backoff may perform very near to ideal scheduling. The accuracy of our modeling approach is verified by simulation in different conditions.
\end{abstract}

\ifCLASSOPTIONdraftclsnofoot
\vspace{-0.5cm}
\fi
\begin{IEEEkeywords}
Exponential backoff, matrix analytic method, random access, slotted Aloha, stability region.
\end{IEEEkeywords}

%
\IEEEpeerreviewmaketitle

\ifCLASSOPTIONdraftclsnofoot
\vspace{-0.3cm}
\fi
\section{Introduction}

\IEEEPARstart{W}{ireless} nodes in a distributed network with a common transmission channel, should independently decide to transmit. The role of a random access MAC protocol is to establish a set of rules that nodes follow in order to avoid collisions during access to the channel. One of the key factors for comparing the performance of different random access MAC protocols is their stability region which is defined as the set of all arrival rate vectors for which the queues in the network are stable, i.e., the property of balanced input-output rates holds. The characterization of the stability region is known to be a very challenging problem. The difficulty lies in the interaction among the queues, i.e., the service process of a queue depends on the status of the other queues \cite{R1}.

Slotted Aloha \cite{R2} is one of the simplest versions of random access MAC protocols, where each node has a fixed transmission probability for transmitting its own packets. Exponential backoff (EB) mechanism \cite{R3} in which nodes dynamically adjust their transmission probabilities to the contention intensity in the network, may improve the protocol to achieve higher throughput. In this mechanism, each node decreases its transmission probability (down to a certain minimum) upon its transmission attempt failure, and resets it upon a successful transmission. The stability region of the slotted Aloha protocol with and without exponential backoff mechanism, has not been clearly understood yet \cite{R4}.

Initial attempts on stability analysis of such networks were for a slotted Aloha network without exponential backoff. In 1979, Tsybakov and Mikhailov \cite{R5} found sufficient conditions for the stability of the queues in the network and found the stability region for two nodes explicitly. In \cite{R1} Rao and Ephremides used the principle of stochastic dominance to find the stability region for a two-node network. The main idea of this technique is to consider dummy packets for some nodes to build networks where some of the buffers are always backlogged (i.e., always have packets). In \cite{R1} authors expressed that when the network has more than two nodes this technique is insufficient to yield the stability region and it can lead to achieve only some inner bounds that are not so tight. The concept of dominant system was used in \cite{R28,R23,R27}, to obtain tight inner and outer bounds to the stability region when the network has more than two nodes.

For three-node slotted Aloha networks with Bernoulli arrival processes, the stability region is characterized in \cite{R19} in which the stability region was described as a function of the stationary probabilities of joint statistics of the queues. Finding these probabilities for networks with more than three nodes, are extremely hard. Moreover, these probabilities could depend on characteristics of arrival processes and are unknown for arbitrary arrival processes. The authors in \cite{R29,R20} generalized the results of \cite{R19} to more general systems of interacting queues. In \cite{R6} an approximate stability region was obtained for an arbitrary number of nodes based on the mean-field asymptotics. The authors claimed that this approximate stability region is exact when the number of nodes grows large and it is accurate for small-sized networks. It is important to note that the measure of accuracy used in \cite{R6} is the number of points where the boundaries of the proposed approximate stability region and the boundaries of exact stability region (found by simulation) intersect. So, irrespective of the claimed accuracy, in some points the obtained stability region may be too far from the exact one.

Exponential backoff leads to memory in the channel uses, because the success probability at consecutive failed transmissions will be different. This stimulates the interaction among nodes and makes the analysis more complex. Moreover, the approach in \cite{R1}, i.e., considering dummy packets to saturate some of the nodes, is not sufficient to find the stability region in such networks. Because, due to memory, the transmission attempts of saturated nodes are not memoryless anymore, hence do not see the other nodes in the steady state, necessarily. 

There are some works that studied the stability of networks with exponential backoff under different assumptions. In order to avoid the inherent difficulty in analyzing the interaction among queues in such networks, most of them considered effects of the exponential backoff mechanism in a scenario similar to Abramson’s model \cite{R2} in which the network has an infinite number of backlogged nodes. This model is in fact a simplified version of the slotted Aloha network. Such simplifications are often used to make analysis more tractable. However, the reported works led to different results which are due to differences in the simplified models. In this respect, it was proven in \cite{R7} that binary exponential backoff is unstable with an infinite number of nodes. However, it was shown that binary exponential backoff can be stable under a finite-node model, if the aggregate arrival rate is sufficiently small \cite{R8}. Since then, different upper bounds for the aggregate arrival rate have been developed \cite{R9,R10} albeit without a common consensus \cite{R11}.

Authors in \cite{R21}, tried to analyze the performance of a multi-packet reception (MPR) slotted Aloha network with exponential backoff mechanism. They assumed that interactions among nodes are statistically independent and found an approximate lower bound for the performance. However, obtained results show poor agreement between analysis and simulations. Also in \cite {R24}, the performance of exponential backoff mechanism over MPR channels was studied, but this analysis is for saturated traffic condition where every node always has a packet to transmit.

In \cite{R4} and \cite{R12}, the authors released some of the aforementioned simplifications and tried to study the stability of the $K$-exponential backoff protocol in a symmetric buffered slotted Aloha network, where $K$ is the cutoff stage of the backoff process. They modeled this network as a multi-queue single-server system and demonstrated that this network with exponential backoff can be stabilized if the backoff factor is properly selected. They used some important simplifying assumptions which make the analysis tractable. They assumed that each packet sees the network in the steady state and always has a success probability equal to its steady state value. Then, they derived this probability based on the independence assumption. It is also noted that, when the number of nodes is small, the correlation among queues is significantly strong and these assumptions result in non-negligible errors \cite{R12}.

In this paper, we focus on determining the stability region of a small-sized buffered Aloha network with exponential backoff mechanism. Although by considering Bernoulli arrival processes, the whole network can be considered as a Markov chain, due to interaction among queues, it cannot be solved by standard methods. However, we model the network with inter-related distributed quasi-birth-death (QBD) processes and include the interaction among queues in the phases of the QBDs. This enables us to discuss about its stability condition with the help of some theorems on the stability of QBDs \cite{R13}. We are also able to evaluate the effect of different backoff factors on the stability region. In our evaluation we use two performance metrics; the $n$-dimensional volume of the stability region and the sum saturation throughput of the network.  Although our modeling approach is an approximate method, our simulations indicate its high accuracy. It is worth noting that unlike most of the existing models \cite{R4,R12,R23,R27}, which deal with very specific packet arrival processes (e.g., Bernoulli), our model is able to accommodate more general packet arrival processes, i.e., discrete-time Markovian arrival processes (D-MAP). 
Succinctly, our contributions in the paper are listed as in the following:

1) We propose a new approach in modeling the interaction among queues in order to be able to handle the model by standard matrix-analytic methods.

2) We derive the stability region of a slotted Aloha network with K-exponential backoff and Bernoulli arrival processes with high accuracy. For the case of two nodes, our region is exact.

3)  We extend our modeling approach in order to cosider D-MAPs. We show that the stability region of a slotted Aloha network with D-MAPs is the same as the stability region of a similar network in which nodes have Bernoulli arrival processes with equivalent average arrival rates.

4) We show that exponential backoff is able to increase the $n$-dimensional volume of the stability region significantly and upgrade the sum saturation throughput of a slotted Aloha network towards an ideal scheduling. 

The remainder of this paper is organized as follows. In Section II, we describe network scenario in more detail. In Section III an approximate model for the network with Bernoulli arrival processes is proposed. In Section IV, based on the proposed model we find the stability region of the network with Bernoulli arrival processes. Although the obtained region is approximated in general, we show that in a two-node network without exponential backoff it intersects with exact region. Moreover, in this section we discussed about the computational complexity of our proposed method. In Section V, we adjust our model to represent the main network with D-MAPs. In Section VI, by doing simulations the accuracy of the obtained stability region is shown. In this section we also show the effect of exponential backoff on the stability region of the slotted Aloha network in different conditions. We conclude the paper in Section VII.

\section{Network Scenario}

In this paper, we consider a small size communication network with $n$ nodes $\lbrace N_1, N_2, \ldots, N_n \rbrace$, where nodes are within the transmission range of each other and share a common channel in a distributed manner. Time is slotted and all packets are assumed to be similar such that transmission time of each packet equals a time slot. Each node has a buffer of infinite capacity to store incoming packets until they are successfully transmitted. It is assumed that each node becomes aware of the status of its transmissions via prompt acknowledgement (ACK) messages.

\subsubsection{The arrival process}
New packets arrive at each node corresponding to an independent D-MAP, which was introduced in \cite{R15,R16}. D-MAP is a nearly general arrival process which can represent a variety of arrival processes which include, as special cases, the Bernoulli arrival process, discrete-time phase-type (PH) renewal process, and Markov modulated Bernoulli process (MMBP). Formally, in D-MAP the arrivals are governed by an underlying discrete-time Markov chain having probability $d_0(u,v)$ with a transition from state $u$ to $v$ without an arrival and having probability $d_1(u,v)$ with a transition from state $u$ to $v$  with an arrival.

Let us define $c_i$ as the number of states of the arrival Markov chain corresponding to node $N_i$. So, the matrix $D_{0,i}$ with elements $d_{0,i}(u,v),  1 \leq u,v \leq c_i$, governs transitions without arrivals, while the matrix $D_{1,i}$ with elements $d_{1,i}(u,v) ,  1 \leq u,v \leq c_i$, governs transitions corresponding to an arrival. Let $\lambda_{i}(u)=\sum_{v}{d_{1,i}(u,v)}$ be the arrival probability for node $N_i$ when it is in state $u$. So, if $\pi^{A}_i(u)$ denotes the stationary probability of being in state $u$, the average arrival rate of packets at node $ N_i $ is $\lambda_i=\sum_{u}{\pi^{A}_i(u) \lambda_{i}(u)}$. It is worth noting that in this paper we assume that arrival Markov chains of D-MAPs are irreducible. We make this assumption for simplifying the presentation, because for D-MAPs with reducible arrival Markov chains, the arguments become complex and it is beyond the scope of this paper.

\subsubsection{The service process}
In order to access the channel, nodes contend with each other based on slotted Aloha with $K$-exponential backoff protocol. Thus, at each time slot that a node $(N_i)$ has a data packet, it attempts to transmit its head-of-line (HOL) packet with a probability. At its first transmission, the packet is transmitted with probability $p_i$. Every time a transmitted packet is collided, the transmission probabilities of the colliding nodes are divided by the backoff factor $r$ $(r\geq1)$. The dividing process of transmission probabilities continues up to $K$ stages. Then, transmission attempts will continue with the probability corresponding to $K^{th}$ stage. If the transmission is successful, the transmission probability of the corresponding node $(N_i)$ will be reset to its initial value, $p_i$. So, a packet for node $N_i$ which experienced $b_i$ number of collisions has a transmission probability of ${p_i}/{r^{b_i}}$; $b_i=0,1,\ldots,K$, where $b_i$ and $K$ are called as the backoff stage of node $N_i$ and the cutoff stage, respectively \cite{R4}. Obviously the slotted Aloha without exponential backoff can be considered as a special case of this scenario corresponding to $K=0$. It is worth noting that for the sake of simplicity, we ignore the physical layer effects of the channel and capture effect. So, any simultaneous transmissions lead to a collision.


In our network scenario the arrival processes of nodes are considered to be D-MAP and the initial transmission probabilities are considered to be asymmetric. Therefore, it represents a nearly general form of slotted Aloha network with exponential backoff. It is also worth noting that the Bernoulli process is a simple example of this general form of arrival processes, i.e., when $c_i=1$, $D_{0,i}=1-\lambda_i$ and $D_{1,i}=\lambda_i$ for each $i \in \lbrace 1,\ldots,n \rbrace$. Due to its simplicity, in Sections III and IV, we propose an approximate model for a network with Bernoulli arrival processes and find its corresponding stability region. Our discussion in Sections III and IV, sheds light onto the stability region of a network with D-MAPs that is discussed in Section V.

\section{Analytical Model for the Network with Bernoulli Arrival Process}

The exact mathematical model for the network described in Section II, with Bernoulli arrival processes, consists of a multi-dimensional Markov chain. The states of this Markov chain are represented by $((q_1,b_1),(q_2,b_2),\ldots,(q_n,b_n))$, where $q_i$ and $b_i$ denote the queue size and backoff stage of the $i^{th}$ node, respectively. The transition probabilities for this chain can be obtained explicitly, but what makes the analysis difficult is due to involving interacting queues. In this scenario, at a time slot the packet transmission of a node can be successful only if none of the other nodes transmits at that slot. So, the success probability of a node's transmission depends on the fact that the queues of other nodes are empty or not. It leads to different transition probabilities on the boundaries of this multi-dimensional Markov chain (i.e., the states that some of nodes are empty) from the ones in the interior of the state space.  Moreover, the number of states is infinite. Thus, traditional analysis methods are inadequate to solve this Markov chain.

Indeed, due to hardness in finding the exact solution for slotted Aloha systems, it seems that resorting to approximate approaches is inevitable. In this respect, a general approach is truncation of the state space in a way that the number of states becomes finite. But by truncating the state space and considering only a finite number of states, our scenario will be changed to a network that nodes have a finite buffer size. So, increasing arrival rates cannot make nodes unstable. Thus, determining the stability region is meaningless. Therefore, we need to consider a new approximate technique to obtain the stability region.

The stability region comprises all arrival rates that keep the network stable. We said a network is stable if all nodes satisfy the condition of balanced input-output rates \cite{R4}. In some networks this notion is equivalent to delay-limited stability region. The latter region is included in the former region but it is not specified except for some special cases \cite{R4}. We focus on the former region throughout the paper. To determine the stability region, instead of using the exact multi-dimensional Markov chain, we model the network approximately with $n$ inter-related Markov chains $({MC}_1,{MC}_2,\ldots,{MC}_n)$. Each ${MC}_i$ represents the status of the $i^{th}$ node ($N_i$), including its buffer status, backoff stages of all nodes, and an indication of the other factors which are effective in the transmission process of node $N_i$. Our aim is to do modeling in a way that each Markov chain is a homogenous quasi-birth-death (QBD) process with a finite number of phases, so enables us to use theorems on the stability condition of homogenous QBD processes \cite{R13}.

In general, a QBD process is a Markov chain comprised of states $\{ (\ell,i) \mid \ell \geq 0, 1\leq i \leq m_\ell\}$, where the state space can be divided into levels, and each level $\ell$ has $m_\ell$ states (phases). In a QBD process, transitions are allowed only to the neighboring levels or within the same level. When transition probabilities between levels (except the first level, $\ell=0$) are alike, the QBD is said to be \textit{homogeneous}. Thus, a homogenous QBD process has a transition probability matrix of the form \cite{R13}:
\begin{equation}\label{2}
P =
\ifCLASSOPTIONdraftclsnofoot
\renewcommand{\arraystretch}{0.6}
\fi
 \left( \begin{array}{ccccc} 
		B_1  & B_0   \\
		B_2 & A_1 & A_0  \\
		 & A_2 & A_1 & A_0 \\
		 & & \ddots & \ddots & \ddots \end{array}  \right),
\end{equation}
where $A_0$, $A_1$, and $A_2$ are square matrices of order $m$, $B_1$ is a square matrix of order $m_0$, and $B_0$ and $B_2$ are rectangular matrices of order $m_0 \times m$ and $m \times m_0$, respectively.

Before we present the state space of each ${MC}_i$, we describe here the notation that we will use throughout the paper. We use bold fonts to represent vectors as opposed to scalars. The elements of a vector are represented by a subscript on the vector symbol. Let $\boldsymbol{q}=(q_1,\ldots,q_n)$ be the queue length vector, in which the queue length of each node $N_i$ is $q_i$. Let $\hat{q}_i$ be the status indicator of $q_i$ which is defined as 
\begin{equation}\label{3}
\hat{q_i} = I(q_i>0),
\end{equation}
where $I(.)$ is the indicator function. Then, the notation $\hat{\boldsymbol{q}}_{-i}$ is used to represent vector $(\hat{q}_{1},\allowbreak \ldots,\allowbreak \hat{q}_{i-1},\allowbreak \hat{q}_{i+1},\allowbreak \ldots,\allowbreak \hat{q}_{n})$ in which status indicators of all queue lengths except $q_i$ are specified. (i.e., subscript `$-i$' on $\hat{\boldsymbol{q}}$ stands for `except node $N_i$'.)

Using these notations, let us define $S_i=\lbrace (q_i,(\boldsymbol{b}_i ,\hat{\boldsymbol{q}}_{-i})) \rbrace$ as the set of states for ${MC}_i$, $i=1,\ldots,n$, where $\boldsymbol{b}_i$ denotes the backoff stage vector and is of the form $\boldsymbol{b}_i=(b_i,{\boldsymbol{b}}_{-i})$. As mentioned before, ${\boldsymbol{b}}_{-i}$ is a vector that represents the backoff stages of all nodes except $N_i$. The first coordinate of states, $q_i$, is called the level of ${MC}_i$, and the second coordinate, $(\boldsymbol{b}_i,\hat{\boldsymbol{q}}_{-i})$, is called the phase of the state $s=(q_i,(\boldsymbol{b}_i,\hat{\boldsymbol{q}}_{-i}))\in S_i$. The level also denotes the whole subset of states with the same first coordinate. 

In this Markov chain one-step transitions are restricted to states in the same or the adjacent levels. It is the result of the fact that in a single time slot the number of packets of $N_i$ can increase one unit according to an arrival (Bernoulli process), decrease one unit according to a successful transmission, or remain the same. So, we have a discrete time QBD process. Moreover, only emptiness or non-emptiness of the queue of $N_i$ is important in the transition probabilities of ${MC}_i$ and the exact number of packets is not effective. So, our QBD process is a homogenous one with the transition probability matrix ($P$) of the form (\ref{2}).

When the queue of node $N_j$ has some packets, so $\hat{q}_j=1$, its backoff stage ($b_j$) can be any integer between $0$ and $K$. But when the queue is empty, its backoff stage is always $0$. So, $(\boldsymbol{b}_{-i},\hat{\boldsymbol{q}}_{-i})$ takes ${(K+2)}^{n-1}$ different values. In level $0$ the queue of node $N_i$ is empty which leads to $b_i=0$. Thus, the number of phases in level $0$ is $m_0={(K+2)}^{n-1}$. In other levels $b_i$ takes $K+1$ different values ($0,\ldots,K$). So, the number of phases in level $l$ when $l \geq 1$ is $m=m_\ell=(K+1){(K+2)}^{n-1}$. If we line up phases of each level in lexicographic order we can define $\varphi_i (\ell,h),h=1,\ldots,m_\ell$ as the $h^{th}$ phase of $\ell^{th}$ level of ${MC}_i$. Thus, ${MC}_i$ can be shown as in Fig. 1.

\begin{figure}
\centering

\ifCLASSOPTIONdraftclsnofoot
\includegraphics[width=3.5in]{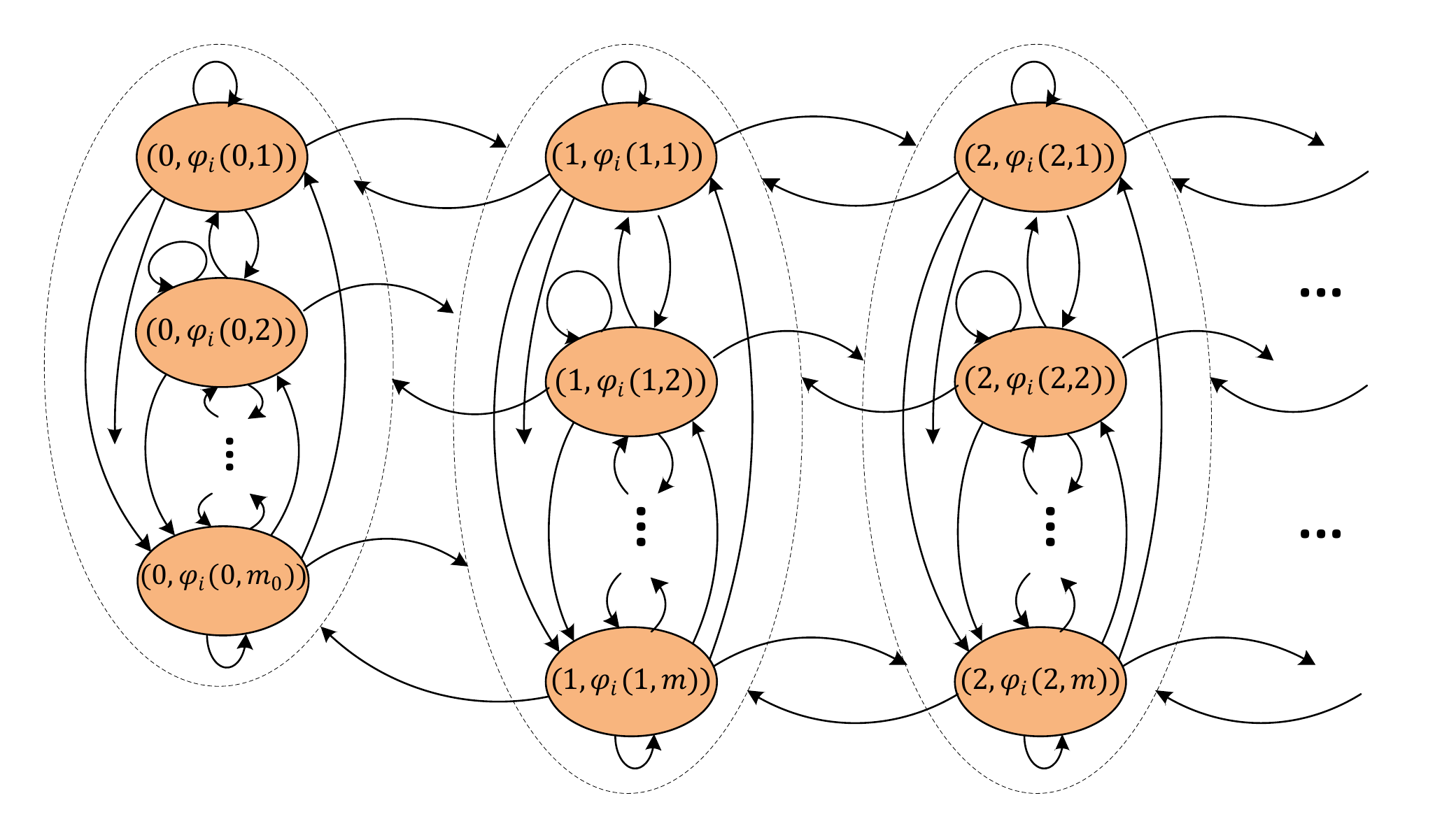}
\vspace{-0.7cm}
\else
\includegraphics[width=0.9\columnwidth,height=1.5in]{Drawing_Exp.pdf}
\vspace{-0.3cm}
\fi

\caption{The Markov chain of the $i^{th}$ node (${MC}_i$).} 
\ifCLASSOPTIONdraftclsnofoot
\vspace{-0.7cm}
\else
\vspace{-0.3cm}
\fi
\label{Markov}
\end{figure} 

In order to find inner blocks of transition probability matrix (i.e., $A_k$ and $B_k$ for $k \in \lbrace 0,1,2 \rbrace$), we should find transition probabilities among the states of ${MC}_i$. Let us denote the transition probability from the current state $(q_i,(\boldsymbol{b}_i,\hat{\boldsymbol{q}}_{-i}))$ to the next state $(q'_i,(\boldsymbol{b}'_i,{\hat{\boldsymbol{q}}_{-i}}'))$ by $P_{(q_i,(\boldsymbol{b}_i,\hat{\boldsymbol{q}}_{-i}))\rightarrow(q'_i,(\boldsymbol{b}'_i,{\hat{\boldsymbol{q}}_{-i}}'))}$. It can be derived as
\ifCLASSOPTIONdraftclsnofoot
\begin{equation}\label{4}
P_{(q_i,(\boldsymbol{b}_i,\hat{\boldsymbol{q}}_{-i}))\rightarrow(q'_i,(\boldsymbol{b}'_i,{\hat{\boldsymbol{q}}_{-i}}'))}= 
\sum_{\substack{\boldsymbol{a}=(a_i,\boldsymbol{a}_{-i}): \\ a_k \in \lbrace 0,1 \rbrace}} [P^{tr}(\boldsymbol{a} \mid s) I(f^{B}(\boldsymbol{b}_i,\boldsymbol{a})=\boldsymbol{b}'_i)
P^{Q}(q'_i,{\hat{\boldsymbol{q}}_{-i}}' \mid q_i,\hat{\boldsymbol{q}}_{-i},\boldsymbol{a})],
\end{equation}
\else
\small
\begin{multline}\label{4}
P_{(q_i,(\boldsymbol{b}_i,\hat{\boldsymbol{q}}_{-i}))\rightarrow(q'_i,(\boldsymbol{b}'_i,{\hat{\boldsymbol{q}}_{-i}}'))}= \\ 
\sum_{\substack{\boldsymbol{a}=(a_i,\boldsymbol{a}_{-i}): \\ a_k \in \lbrace 0,1 \rbrace}} P^{tr}(\boldsymbol{a} \mid s) I(f^{B}(\boldsymbol{b}_i,\boldsymbol{a})=\boldsymbol{b}'_i)
P^{Q}(q'_i,{\hat{\boldsymbol{q}}_{-i}}' \mid q_i,\hat{\boldsymbol{q}}_{-i},\boldsymbol{a}),
\end{multline}
\normalsize
\fi
where $P^{tr}(\boldsymbol{a} \mid s)$ denotes the probability that the transmission attempt vector $\boldsymbol{a}=(a_i,\boldsymbol{a}_{-i})$ is made by the nodes when the current state is $s=(q_i,(\boldsymbol{b}_i,\hat{\boldsymbol{q}}_{-i}))$. The transmission attempt vector $\boldsymbol{a}$ is an $n$-component vector that its $k^{th}$ component ($k \in \lbrace 1,\ldots,n \rbrace $) specifies whether node $N_k$ is attempting to transmit a packet in a corresponding time slot (i.e., $a_k=1$) or not (i.e., $a_k=0$). Notation $f^{B}(\boldsymbol{b}_i,\boldsymbol{a})$ denotes the backoff function that determines the backoff stage vector in the next slot when the current backoff stage vector is $\boldsymbol{b}_i$ and the transmission attempt vector is $\boldsymbol{a}$. Also, notation $P^{Q}(q'_i,{\hat{\boldsymbol{q}}_{-i}}' \mid q_i,\hat{\boldsymbol{q}}_{-i},\boldsymbol{a})$ denotes a conditional probability function that determines the probability of changing the status of queues to $(q'_i,{\hat{\boldsymbol{q}}_{-i}}')$ when previous status of queues is $(q_i,\hat{\boldsymbol{q}}_{-i})$ and the transmission attempt vector is $\boldsymbol{a}$.

$P^{tr}(\boldsymbol{a} \mid s)$ can be obtained as
\begin{equation}\label{5}
P^{tr}(\boldsymbol{a} \mid s)=\prod_{k=1}^{n} {P_k^{tr} (a_k \mid b_k,\hat{q}_{k})} ,
\end{equation}
where $P_k^{tr} (a_k \mid b_k,\hat{q}_{k})$ is the probability for the attempt of $N_k$ ($a_k=1$ for transmission) where its backoff stage is $b_k$ and its status indicator is $\hat{q}_{k}$. Thus,
\ifCLASSOPTIONdraftclsnofoot
\begin{equation}\label{6}
P_k^{tr} (1 \mid b_k,\hat{q}_{k})=
1-P_k^{tr} (0 \mid b_k,\hat{q}_{k})=\dfrac {p_k} {r^{b_k}} I(\hat{q}_{k}=1); k=1,\ldots,n . 
\end{equation}
\else
\begin{multline}\label{6}
P_k^{tr} (1 \mid b_k,\hat{q}_{k})=\\
1-P_k^{tr} (0 \mid b_k,\hat{q}_{k})=\dfrac {p_k} {r^{b_k}} I(\hat{q}_{k}=1); k=1,\ldots,n . 
\end{multline}
\fi
The backoff function $f^{B}(\boldsymbol{b}_i,\boldsymbol{a})$, is obtained by
\begin{equation}\label{7}
f^{B}(\boldsymbol{b}_i,\boldsymbol{a})=(f_i^{B}(b_i,\boldsymbol{a}),f_{-i}^{B}(\boldsymbol{b}_{-i},\boldsymbol{a})),
\end{equation}
where
\ifCLASSOPTIONdraftclsnofoot
\begin{equation}\label{8}
f_{-i}^{B}(\boldsymbol{b}_{-i},\boldsymbol{a})=
(f_1^{B}(b_1,\boldsymbol{a}),\ldots,f_{i-1}^{B}(b_{i-1},\boldsymbol{a}),f_{i+1}^{B}(b_{i+1},\boldsymbol{a}),\ldots,f_n^{B}(b_n,\boldsymbol{a})),
\end{equation}
\else
\small
\begin{multline}\label{8}
f_{-i}^{B}(\boldsymbol{b}_{-i},\boldsymbol{a})=\\
(f_1^{B}(b_1,\boldsymbol{a}),\ldots,f_{i-1}^{B}(b_{i-1},\boldsymbol{a}),f_{i+1}^{B}(b_{i+1},\boldsymbol{a}),\ldots,f_n^{B}(b_n,\boldsymbol{a})),
\end{multline}
\normalsize
\fi
and $f_k^{B}(b_k,\boldsymbol{a})$ denotes the next backoff stage of node $N_k$ as follows:
\begin{equation}\label{9}
f_k^{B}(b_k,\boldsymbol{a})=
\left\{
	\begin{array}{lll} 
		b_k & \mbox{if } a_k=0, \\
		0  & \mbox{if } a_k=1, \forall j \neq k : a_j=0,  \\
		\min (b_k+1,K) & \mbox{if } a_k=1, \exists j \neq k : a_j=1.
	\end{array}
\right.
\end{equation}
In fact, in (\ref{9}), $a_k=0$ means that $N_k$ is not attempting to transmit, so, its backoff stage remains unchanged. When $N_k$ is the only node which is attempting to transmit, its transmission is successful and its backoff stage becomes zero. It means that the transmission probability of node $N_k$ resets to its initial value $p_k$. At last, if node $N_k$ as well as at least another node are attempting to transmit in a slot concurrently, a collision occurs. So the backoff stage increases one unit, if it does not exceed the cutoff stage ($K$).

Moreover, $P^{Q}(q'_i,{\hat{\boldsymbol{q}}_{-i}}' \mid q_i,\hat{\boldsymbol{q}}_{-i},\boldsymbol{a})$ is given by
\ifCLASSOPTIONdraftclsnofoot
\begin{equation}\label{10}
P^{Q}(q'_i,{\hat{\boldsymbol{q}}_{-i}}' \mid q_i,\hat{\boldsymbol{q}}_{-i},\boldsymbol{a})=
\sum_{\substack{\boldsymbol{d}=(d_i,\boldsymbol{d}_{-i}): \\ d_k \in \lbrace 0,1 \rbrace, |\boldsymbol{d}|=0,1}} [P_i^D (\boldsymbol{a},\boldsymbol{d}) P_i^{A}(q'_i-q_i+d_i)
\prod_{j \neq i} {P_j^{I}(\hat{q_j},{\hat{q_j}}',{\hat{q_j}}'-\hat{q_j}+d_j)}].
\end{equation}
\else
\begin{multline}\label{10}
P^{Q}(q'_i,{\hat{\boldsymbol{q}}_{-i}}' \mid q_i,\hat{\boldsymbol{q}}_{-i},\boldsymbol{a})= \\
\sum_{\substack{\boldsymbol{d}=(d_i,\boldsymbol{d}_{-i}): \\ d_k \in \lbrace 0,1 \rbrace, |\boldsymbol{d}|=0,1}} [P_i^D (\boldsymbol{a},\boldsymbol{d}) P_i^{A}(q'_i-q_i+d_i) \\
\prod_{j \neq i} {P_j^{I}(\hat{q_j},{\hat{q_j}}',{\hat{q_j}}'-\hat{q_j}+d_j)}].
\end{multline}
\fi
The notation $P_i^D(\boldsymbol{a},\boldsymbol{d})$ denotes the probability that the elements of the state which are related to queue lengths, i.e., $\boldsymbol{Q}=(q_i,\hat{\boldsymbol{q}}_{-i})$, decrease to $\boldsymbol{Q}-\boldsymbol{d}$, according to possible departures and before considering the new arrivals, where $\boldsymbol{d}=(d_i,\boldsymbol{d}_{-i})$ is a decrement vector. Also, $P_i^{A} (k)$ and $P_j^{I} (\hat{q_j},{\hat{q_j}}',{\hat{q_j}}'-\hat{q_j}+d_j)$, respectively, denote the probability of arriving $k$ packets for $N_i$ and the probability that the status indicator of $q_j$ (i.e., $\hat{q_j}$) changes from $\hat{q_j}$ to ${\hat{q_j}}'$ as a result of new arrivals for $N_j$ (it needs ${\hat{q_j}}'-\hat{q_j}+d_j$ new packets). Moreover, the notation $|\boldsymbol{d}|$ stands for the $\ell_1$-norm of vector $\boldsymbol{d}$ which is the sum of the absolute values of its components. So, $|\boldsymbol{d}|=0,1$ refers to the fact that at each time slot at most one packet can be transmitted successfully and as a result, at most one of the queue lengths can decrease. It is worth noting that although different arrangements of arrival and service processes in discrete time are possible \cite{R18}, in this paper we consider early arrival model, i.e., service completions occur just before the slot boundaries and the new arrivals come just after the slot boundaries. However, it is very simple to extend the discussions and equations of this paper to other arrangements.


By knowing the exact queue length of each node and assuming a fixed attempt vector, the decrement in queue length of each node is obtained easily. But in order to make the number of phases finite, only an indicator for the queue length of each node $N_j$ ($j \neq i$) is considered in the states of ${MC}_i$ that determines the emptiness or non-emptiness of its corresponding queue. This makes the computation of $P_i^D(\boldsymbol{a},\boldsymbol{d})$ troublesome. Note that $\hat{q}_{j}=1$ contains all $q_j$s that are greater than or equal to one, but among such $q_j$s, $q_j=1$ is different from the others. If the current state corresponds to $q_j=1$, due to a departure it transits to a state corresponding to $\hat{q}_{j}=q_j=0$, but when the current state corresponds to $q_j>1$ its indicator does not change anyway ($\hat{q}_{j}=1$). So, although in defining the states of ${MC}_i$ we do not distinguish between $q_j=1$ and $q_j>1$ (in order to reduce the number of phases), there is a considerable difference between them. Thus, in order to compute $P_i^D(\boldsymbol{a},\boldsymbol{d})$, we need to know the conditional probability of having exactly one packet in the buffer of $N_j$ ($j \neq i$), when $\hat{q}_{j}=1$. So, let us define these conditional probabilities as $z_j = P(q_j=1 \mid \hat{q}_j=1)$, for all $j \neq i$. Now, by using these new parameters, $P_i^D(\boldsymbol{a},\boldsymbol{d})$ can be obtained as
\ifCLASSOPTIONdraftclsnofoot
\begin{equation}\label{11}
P_i^D(\boldsymbol{a},\boldsymbol{d})=
\left\{
	\begin{array}{llll} 
		I(\boldsymbol{d}=\boldsymbol{0}) & \mbox{if } |\boldsymbol{a}|=0 $ or $ |\boldsymbol{a}| > 1, \\
		I(\boldsymbol{d}=(1,\boldsymbol{0})) & \mbox{if } |\boldsymbol{a}|=1 , a_i=1, \\
		z_j I(|\boldsymbol{d}|=1 , d_j=1) \\
		+ (1-z_j) I(\boldsymbol{d}=\boldsymbol{0}) & \mbox{if } |\boldsymbol{a}|=1 , a_j=1 , j \neq i.
	\end{array}
\right.
\end{equation}
\else
\vspace{-0.3cm}
\begin{multline}\label{11}
P_i^D(\boldsymbol{a},\boldsymbol{d})=\\
\left\{
	\begin{array}{llll} 
		I(\boldsymbol{d}=\boldsymbol{0}) & \mbox{if } |\boldsymbol{a}|=0 $ or $ |\boldsymbol{a}| > 1, \\
		I(\boldsymbol{d}=(1,\boldsymbol{0})) & \mbox{if } |\boldsymbol{a}|=1 , a_i=1, \\
		z_j I(|\boldsymbol{d}|=1 , d_j=1) \\
		+ (1-z_j) I(\boldsymbol{d}=\boldsymbol{0}) & \mbox{if } |\boldsymbol{a}|=1 , a_j=1 , j \neq i.
	\end{array}
\right.
\end{multline}
\fi
In fact, in (\ref{11}), $|\boldsymbol{a}|=0$ means that none of the nodes is transmitting. Obviously in this case no departure and no decrement occurs. Moreover, when more than one node are attempting to transmit in the same time slot ($|\boldsymbol{a}| > 1$), due to the collision no decrement occurs, too. Thus, the queue lengths can be decreased only when exactly one of the nodes is attempting to transmit its HOL packet. If this unique transmitter node is $N_i$, its corresponding queue length decreases one unit and the queue lengths of the others do not change. So the decrement vector is of the form $\boldsymbol{d}=(1,\boldsymbol{0})$. At last, when the unique transmitter node is a node except $N_i$ (e.g., $N_j$), HOL packet of its queue departs. If the queue of $N_j$ had exactly one packet, this departure makes it empty. So its queue length indicator decreases one unit ($d_{j}=1$) and changes to $\hat{q}_j=0$. Otherwise, if its queue had more than one packet, this departure preserves its status and the corresponding queue length indicator does not change ($d_{j}=0$). These two situations take place with probabilities $z_j$ and $1-z_j$, respectively.

Moreover, $P_i^{A} (k)$ is obtained by
\begin{equation}\label{12}
P_i^{A} (k)=(1-\lambda_i)I(k=0)+\lambda_i I(k=1).
\end{equation}
That is, due to Bernoulli arrival process, at each time slot at most one packet can arrive at $N_i$ with probability $\lambda_i$.

Also, $P_j^{I} (\hat{q_j},{\hat{q_j}}',k)$ is obtained by
\ifCLASSOPTIONdraftclsnofoot
\begin{equation}\label{13}
P_j^{I} (\hat{q_j},{\hat{q_j}}',k)=
\left\{
	\begin{array}{llll}
	(1-\lambda_j)I(k=0) & \mbox{if } (\hat{q_j},{\hat{q_j}}')=(0,0) $ or $ (1,0),  \\
	\lambda_j I(k=1) & \mbox{if } (\hat{q_j},{\hat{q_j}}')=(0,1), \\
	I(k=0)+\lambda_j I(k=1) & \mbox{if } (\hat{q_j},{\hat{q_j}}')=(1,1).
	\end{array}
\right.
\end{equation}
\else
\begin{multline}\label{13}
P_j^{I} (\hat{q_j},{\hat{q_j}}',k)=\\
\left\{
	\begin{array}{lll} 
	(1-\lambda_j)I(k=0) & \mbox{if } (\hat{q_j},{\hat{q_j}}')=(0,0) $ or $ (1,0),  \\
	\lambda_j I(k=1) & \mbox{if } (\hat{q_j},{\hat{q_j}}')=(0,1), \\
	I(k=0)+\lambda_j I(k=1) & \mbox{if } (\hat{q_j},{\hat{q_j}}')=(1,1).
	\end{array}
\right.
\end{multline}
\fi
The concept behind this function and $P_i^{A} (k)$ is the same, but there is a different notion in $P_j^{I}$ when $(\hat{q_j},{\hat{q_j}}')=(1,1)$ and $k=0$. In this case $P_j^{I}(1,1,0)$ denotes the probability that node $N_j$ stays non-empty when it was non-empty at the previous time slot and since no departure occur, it needs no new packets. It is obvious that for this event it is not important whether an arrival occurs or not. So, the corresponding probability in (\ref{13}) is 1.

Now after calculating transition probabilities, we can derive the inner submatrices of $P$ in (\ref{2}) as in the following:
\begin{IEEEeqnarray}{L}\label{14}
A_0=P_{(\ell,:)\rightarrow(\ell+1,:) }, A_1=P_{(\ell,:)\rightarrow(\ell,:) };     \ell \geq 1,\nonumber\\
B_0=P_{(0,:)\rightarrow(1,:) } , B_1=P_{(0,:)\rightarrow(0,:)},\nonumber\\
B_2=P_{(1,:)\rightarrow(0,:)}, A_2=P_{(\ell,:)\rightarrow(\ell-1,:)};              \ell \geq 2,
\end{IEEEeqnarray}
where $P_{(\ell,:)\rightarrow(\ell',:)}$ denotes a matrix of order $m_\ell \times m_{\ell'}$ that its ${(h,h')}^{th}$ element is the transition probability from the current state $(\ell,\varphi_i (\ell,h))$ to the next state $(\ell',\varphi_i (\ell',h'))$.

Now we can apply matrix-analytic methods \cite{R13} to solve ${MC}_i$ and find the steady state distribution as follows:
\ifCLASSOPTIONdraftclsnofoot
\begin{equation} \label{41}
\setlength{\arraycolsep}{12pt}
\begin{array}{ll} 
A_0+RA_1+R^{2}A_2=R, &
\boldsymbol{\pi}_i(\ell+1)=\boldsymbol{\pi}_i(1)R^{\ell};     \ell \geq 1,
\end{array}
\end{equation}
\else
\begin{IEEEeqnarray}{L}\label{41}
\boldsymbol{\pi}_i(\ell+1)=\boldsymbol{\pi}_i(1)R^{\ell};     \ell \geq 1,\nonumber\\
A_0+RA_1+R^{2}A_2=R,
\end{IEEEeqnarray}
\fi
where $R$ is a matrix such that, for any $\ell \geq 1$, $R_{ij}$ is the expected number of visits to $j^{th}$ phase of level $\ell+1$ before a return to level $\ell$, given that the process starts in $i^{th}$ phase of level $\ell$. Moreover, the vectors $\boldsymbol{\pi}_i(0)$ and $\boldsymbol{\pi}_i(1)$ are such that
\begin{IEEEeqnarray}{L}\label{42}
\boldsymbol{\pi}_i(0)B_1+\boldsymbol{\pi}_i(1)B_2=\boldsymbol{\pi}_i(0),   \nonumber\\
\boldsymbol{\pi}_i(0)B_0+\boldsymbol{\pi}_i(1)A_1+\boldsymbol{\pi}_i(1)RA_2=\boldsymbol{\pi}_i(1),     \nonumber\\
\boldsymbol{\pi}_i(0)\boldsymbol{1}+\boldsymbol{\pi}_i(1){{(I-R)}^{-1}}\boldsymbol{1}=1,
\end{IEEEeqnarray}
where $\boldsymbol{1}$ is a column vector of $1$s.

Now, it is obvious that $\pi_i(\ell)=\boldsymbol{\pi}_i(\ell)\boldsymbol{1}$ is the stationary probability of being at level $\ell$ of ${MC}_i$. Due to the fact that level $\ell$ of ${MC}_i$ includes all states in which $q_i=\ell$, this distribution is equivalent to the stationary queue length distribution of node $N_i$. So, we consider $\pi_k(.)$ as the stationary distribution of the queue lengths of each node $N_k$ ($k=1,\ldots,n$). Now, by some simple manipulations, $z_j$ which is used in (\ref{11}) can be found in terms of $\pi_j(\ell)$, as in the following:
\ifCLASSOPTIONdraftclsnofoot
\begin{equation}\label{15}
z_j=P(q_j=1 | \hat{q}_j=1)=P(q_j=1 | q_j \geq 1)
=\dfrac {P(q_j=1)} {P(q_j \geq 1)}=\dfrac {\pi_j(1)} {1-\pi_j(0)}.
\end{equation}
\else
\begin{multline}\label{15}
z_j=P(q_j=1 | \hat{q}_j=1)=P(q_j=1 | q_j \geq 1)\\
=\dfrac {P(q_j=1)} {P(q_j \geq 1)}=\dfrac {\pi_j(1)} {1-\pi_j(0)}.
\end{multline}
\fi
In our technique, each $\pi_i(\ell)$ is computed by solving corresponding ${MC}_i$ in which some of the transition probabilities are expressed in terms of $z_j$s ($\forall j \neq i$). Moreover, (\ref{15}) shows that $z_j$s are calculated based on $\pi_j(\ell)$s. So, in order to find stationary queue length distribution of all nodes of the network, Markov chains ${MC}_1,\ldots,{MC}_n$, should be solved recursively.

\section{Stability Region of the Network with Bernoulli Arrival Process}
By using the model proposed in the previous section, we find the stability region of the network with Bernoulli arrival processes. A network is said to be stable if all nodes satisfy the condition of balanced input-output rates. In our model, this condition is equivalent to existence of stationary distributions for all ${MC}_i$s. Due to the fact that in our model all QBDs (${MC}_1,\ldots,{MC}_n$) are irreducible and aperiodic, the stability condition is equivalent to positive recurrency of QBDs. In other words, for a given vector of arrival rates if and only if all QBDs are positive recurrent, we can conclude that this vector is in the stability region. So, in order to determine the stability region, we follow Theorem 1 presented by Latouche and Ramaswami in \cite{R13}.

\newtheorem{theorem}{Theorem}
\begin{theorem}
If a homogenous QBD is irreducible and the number of phases is finite, and if the corresponding stochastic matrix $A=A_0+A_1+A_2$ is irreducible, then the QBD is positive recurrent if and only if $\mu=\boldsymbol{\alpha}(A_0-A_2)\boldsymbol{1}<0$, where $\boldsymbol{\alpha}$ is the stationary probability vector of $A$ and $\boldsymbol{1}$ is a column vector of 1’s.
\end{theorem}

It is clear that in our model each ${MC}_i$ and its corresponding matrix $A$ are irreducible and the number of phases is finite. So, it can be derived from Theorem 1 that the stability region comprises all arrival rates for which the condition $\mu<0$ is satisfied for all QBDs (i.e., ${MC}_1,\ldots,{MC}_n$). Thus, in order to find the stability region, first we should calculate the value of $\mu$ corresponding to each ${MC}_i$ in terms of arrival rates. $\mu$ is comprised of two terms: $\boldsymbol{\alpha}$ which is the stationary probability vector of $A$ and $(A_0-A_2)\boldsymbol{1}$. So for calculating $\mu$, we calculate $A$ and $(A_0-A_2)\boldsymbol{1}$ for ${MC}_i$ in the following.

$A$ is a square matrix of order $m$ that its ${(h,h')}^{th}$ element is obtained by:
\ifCLASSOPTIONdraftclsnofoot
\setlength{\arraycolsep}{0.0em}
\begin{IEEEeqnarray}{L}\label{16}
A(h,h')=A_0(h,h')+A_1(h,h')+A_2(h,h')=   \nonumber\\
P_{(\ell,h)\rightarrow(\ell+1,h')}+P_{(\ell,h)\rightarrow(\ell,h')}+P_{(\ell,h)\rightarrow(\ell-1,h')}=
\sum_{\ell'}{P_{(\ell,h)\rightarrow(\ell',h')}} ,         \ell \geq 2 .
\end{IEEEeqnarray}
\setlength{\arraycolsep}{0pt}
\else
\begin{align}\label{16}
A(h,h')=A_0(h,h')+A_1(h,h')+A_2(h,h')=&\nonumber\\
P_{(\ell,h)\rightarrow(\ell+1,h')}+P_{(\ell,h)\rightarrow(\ell,h')}+P_{(\ell,h)\rightarrow(\ell-1,h')}=&\nonumber\\
\sum_{\ell'}{P_{(\ell,h)\rightarrow(\ell',h')}} ,         \ell \geq 2 .&                           
\end{align}
\fi
Note that the last equality holds because in QBDs $P_{(\ell,h)\rightarrow(\ell',h')}=0$ for $|\ell-\ell'|>1$. Moreover, $\boldsymbol{\alpha}(h)$ is the stationary probability of being at one of the states of $S(h)=\lbrace s=(\ell,\varphi_i(\ell,h)) \mid \ell \geq 2 \rbrace$ when the transition from $S(h)$ to $S(h')$ takes place with probability $A(h,h')$.

\begin{theorem}
In each ${MC}_i$, $(A_0-A_2)\boldsymbol{1}$ is a column vector of order $m$ that its $h^{th}$ element is equal to $\lambda_i-p_{i}^{suc}(h)$, where $p_{i}^{suc}(h)=P^{tr}((1,\boldsymbol{0}) \mid s)$ is the probability that the $i^{th}$ node has a successful transmission when the current state is $s=(\ell,\varphi_i(\ell,h)), \ell \geq 2$. 
\end{theorem}

\begin{IEEEproof}
When we discuss about $A_i$s, it means that the queue of node $N_i$ has at least two packets, i.e., $\ell \geq 2$ (see (\ref{2})). So at each phase there is a probability that the HOL packet of this queue is transmitted successfully, i.e., only $N_i$ attempts to transmit (i.e., $\boldsymbol{a}=(1,\boldsymbol{0})$). We define this probability as $p_{i}^{suc}(h)=P^{tr}((1,\boldsymbol{0}) \mid s)$. So, the $h^{th}$ element of $(A_0-A_2)\boldsymbol{1}$ is found as
\ifCLASSOPTIONdraftclsnofoot
\setlength{\arraycolsep}{0.0em}
\begin{IEEEeqnarray}{L}\label{17}
[(A_0-A_2)\boldsymbol{1}](h)=\sum_{h'}{P_{(\ell,h)\rightarrow(\ell+1,h')}}-\sum_{h'}{P_{(\ell,h)\rightarrow(\ell-1,h')}} =\nonumber\\
(1-p_{i}^{suc}(h))\lambda_i-p_{i}^{suc}(h)(1-\lambda_i )=\lambda_i-p_{i}^{suc}(h).
\end{IEEEeqnarray}
\setlength{\arraycolsep}{0pt}  
\else
\begin{multline}\label{17}
[(A_0-A_2)\boldsymbol{1}](h)=\sum_{h'}{P_{(\ell,h)\rightarrow(\ell+1,h')}}-\sum_{h'}{P_{(\ell,h)\rightarrow(\ell-1,h')}} \\
=(1-p_{i}^{suc}(h))\lambda_i-p_{i}^{suc}(h)(1-\lambda_i )=\lambda_i-p_{i}^{suc}(h).
\end{multline}
\fi
\end{IEEEproof}
Theorem 2 along with definition of $\mu$ results in:
\begin{equation}\label{18}
\mu=\sum\limits_{h=1}^m {\boldsymbol{\alpha}(h)(\lambda_i-p_{i}^{suc}(h))}=\lambda_i-\sum\limits_{h=1}^m {\boldsymbol{\alpha}(h)p_{i}^{suc}(h)}.
\end{equation}
From Theorem 1 and (\ref{18}) we conclude that the necessary and sufficient condition for ${MC}_i$ to be positive recurrent is $\mu=\lambda_i-\sum\nolimits_{h=1}^m {\boldsymbol{\alpha}(h)p_{i}^{suc}(h)}<0$. In the following theorem, we will prove that when node $N_i$ is saturated (i.e., always has packets and the probability of having finite number of packets is zero), $\sum\nolimits_{h=1}^m {\boldsymbol{\alpha}(h)p_{i}^{suc}(h)}$ is independent of $\lambda_i$.

\begin{theorem}
When node $N_i$ is saturated, $\sum\nolimits_{h=1}^m {\boldsymbol{\alpha}(h)p_{i}^{suc}(h)}$ is a function of ${\boldsymbol{\lambda}}_{-i}$ and is independent of $\lambda_i$.
\end{theorem}

\begin{IEEEproof}
By definition it is obvious that $p_{i}^{suc}(h)$ is independent of all arrival rates. So, we discuss only about $\boldsymbol{\alpha}$ which is the stationary probability vector of $A$. It is easily derived from (\ref{16}) that for calculating the elements of $A$, levels of the destined states are not important. Thus, the arrival rate of $N_i$ which only affects the level of the states, does not appear in $A$ directly. So, the elements of $A$ are only functions of ${\boldsymbol{\lambda}}_{-i}$ and $\boldsymbol{z}_{-i}$. $\boldsymbol{z}_{-i}$ can be found by (\ref{15}) when ${\boldsymbol{\pi}}_{j}(\ell)$ is calculated by solving all QBDs except ${MC}_i$ by assuming $N_i$ is saturated ($z_i=0$). 

When we consider $N_i$ is saturated, states of ${MC}_j$ ($j \neq i$) in which $\hat{q}_i=0$ become transient. So, in solving ${MC}_j$ we consider only the states in which $\hat{q}_i=1$. This makes analysis of ${MC}_j$ and as a result $z_j$, independent of $\lambda_i$. As mentioned before, the elements of $A$ are only functions of ${\boldsymbol{\lambda}}_{-i}$ and $\boldsymbol{z}_{-i}$, and now, it is proved that $\boldsymbol{z}_{-i}$ is a function of ${\boldsymbol{\lambda}}_{-i}$ and independent of $\lambda_i$. So, it is concluded that $A$ and as a result $\boldsymbol{\alpha}$, are only functions of ${\boldsymbol{\lambda}}_{-i}$, hence independent of $\lambda_i$.
\end{IEEEproof}

So, for a fixed ${\boldsymbol{\lambda}}_{-i}$ the value of $\lambda_i$ which makes $N_i$ and equivalently ${MC}_i$ be in the border of instability is derived as follows:
\begin{equation}\label{19}
\lambda_{i}^{SR}({\boldsymbol{\lambda}}_{-i}) \equiv \sum\limits_{h=1}^m {\boldsymbol{\alpha}(h)p_{i}^{suc}(h)},
\end{equation}
where the term of right side of (\ref{19}) should be calculated when node $N_i$ is saturated.

It can be concluded from (\ref{19}) that a vector of arrival rates $\boldsymbol{\lambda}=(\lambda_1,\ldots,\lambda_n)$ keeps the network stable if and only if it satisfies $\lambda_i<\lambda_{i}^{SR}({\boldsymbol{\lambda}}_{-i})$ for all $i \in \lbrace 1,\ldots,n \rbrace$. So, the stability region can be expressed as follows:
\begin{equation}\label{20}
SR=\lbrace \boldsymbol{\lambda}=(\lambda_1,\ldots,\lambda_n) \mid \lambda_i<\lambda_{i}^{SR}({\boldsymbol{\lambda}}_{-i}) , \forall i; 1 \leq i \leq n \rbrace .
\end{equation}
Therefore, in order to find the stability region ($SR$), for each $i$, $\lambda_{i}^{SR}({\boldsymbol{\lambda}}_{-i})$ should be calculated as in (\ref{19}) for different values of ${\boldsymbol{\lambda}}_{-i}$. This can be done systematically, by choosing a small enough step size ($\Delta \lambda$) and calculate $\lambda_{i}^{SR}$ for ${\boldsymbol{\lambda}}_{-i}$s in which all $\lambda_j$s ($j \neq i$) are integer multiples of the step size. It is obvious that for the stability of node $N_j$, its arrival rate should be less than or equal to its maximum transmission probability (i.e., $\lambda_j \leq p_j$). So, $\lambda_{i}^{SR}({\boldsymbol{\lambda}}_{-i})$ should be calculated for $\Lambda_i=\lbrace {\boldsymbol{\lambda}}_{-i} \mid \forall j \neq i, \lambda_j=k_j\Delta \lambda, 0 \leq k_j \leq \left\lfloor \frac {p_j} {\Delta \lambda} \right\rfloor \rbrace$. Then, the curve $\lambda_{i}=\lambda_{i}^{SR}({\boldsymbol{\lambda}}_{-i})$ can be approximately derived. The region which is below all these curves is the stability region of the network. All steps to find the stability region are summarized in Fig. 2.

\begin{figure}
\centering

\ifCLASSOPTIONdraftclsnofoot
\includegraphics[width=0.5\columnwidth,height=4.2in]{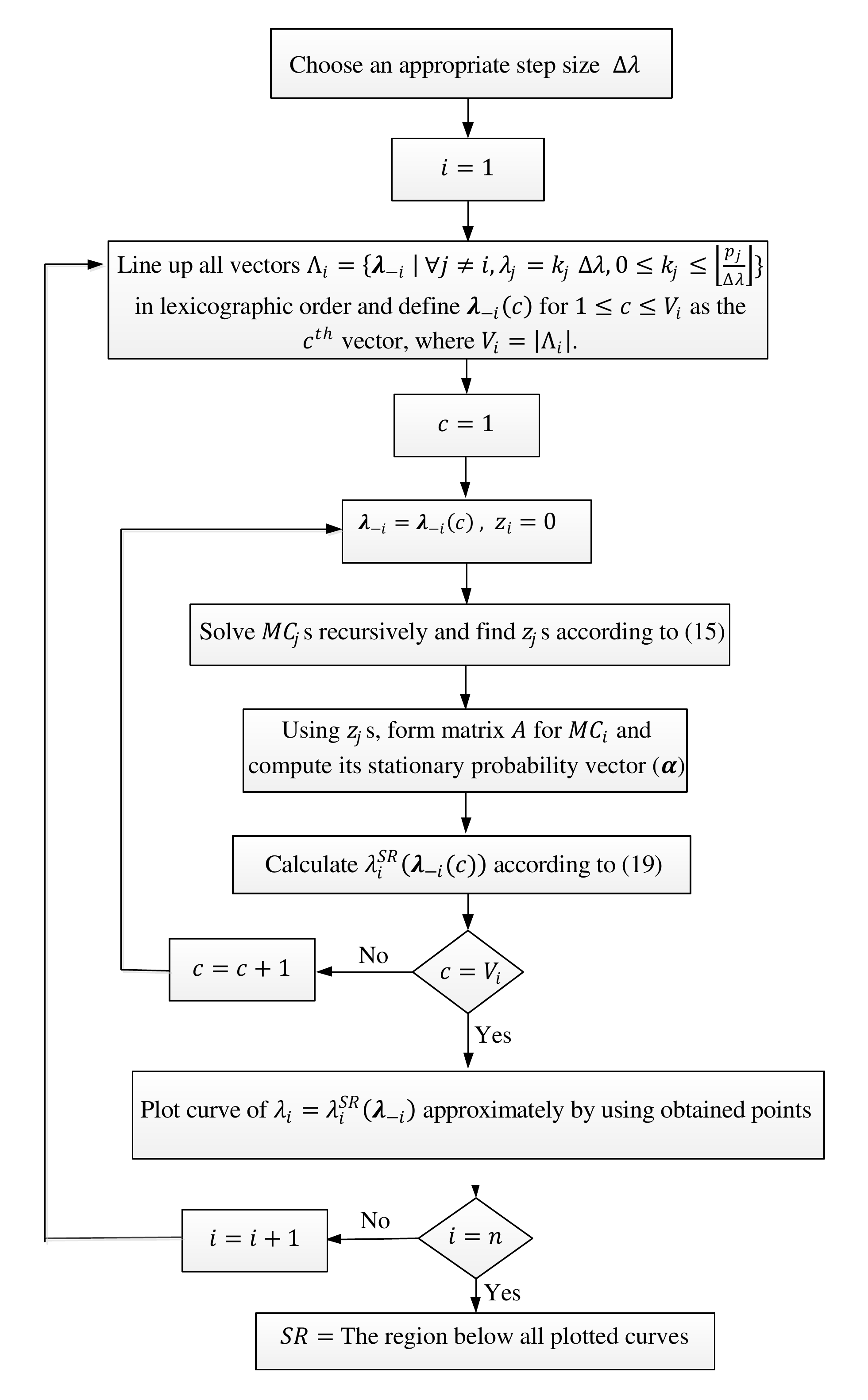}
\vspace{-0.7cm}
\else
\includegraphics[width=2.5in,height=4.2in]{Drawing_Exp,flowchart.pdf}
\vspace{-0.3cm}
\fi

\caption{Flowchart for showing the algorithm of finding the stability region.} 
\ifCLASSOPTIONdraftclsnofoot
\vspace{-0.7cm}
\else
\vspace{-0.3cm}
\fi
\label{Flowchart}
\end{figure} 

As shown in Fig. 2. for a fixed $i$ and $\boldsymbol{\lambda}_{-i}$, $n-1$ QBDs should be solved recursively. It has been shown that solving a homogenous QBD with $m$ phases is of order $O(m^3)$ \cite{R13}. So, solving $n-1$ QBDs each with $(K+1){(K+2)}^{n-1}$ phases is of order $O(n{(K+2)}^{3n})$. Our numerical results show that the convergence of solving QBDs is fast and regardless of size of the network, it usually converges in at most $20$ steps. So, the effect of considering the iteration steps is only a constant factor, which in big-$O$ notation is discarded.

Assume for each $i \in \lbrace 1,\ldots,n \rbrace$, $V_i$ denotes the number of $\boldsymbol{\lambda}_{-i}$s for which $\lambda_{i}^{SR}({\boldsymbol{\lambda}}_{-i})$ should be calculated. It is clear that
\begin {equation}\label{39}
V_i=\prod_{j \neq i} {(\left\lfloor \dfrac {p_j} {\Delta \lambda} \right\rfloor +1)}.
\end{equation}
So, the number of times that QBDs should be solved together is
\begin {equation}\label{40}
\sum_{i=1}^{n} V_i \leq n \max_i {V_i} \leq n {(\dfrac {1} {\Delta \lambda} +1)}^{n-1}.
\end{equation}
Thus, for a fixed step size (i.e., $\Delta \lambda$) the computational complexity of finding stability region for a network with $n$ nodes, is $O(n^2{[{(K+2)}^3( {\Delta \lambda}^{-1}  +1)]}^{n})$. Since the number of operations required by this method grows exponentially with the number of nodes, due to its complexity, it is not suitable to be applied on large-scale networks.

Note that although our proposed model is an approximate approach in general, in the case of a two-node network without exponential backoff ($K=0$), it leads to the exact stability region. As discussed before, the stability region which is found by our method is derived as
\begin{equation}\label{21}
SR=\lbrace (\lambda_1,\lambda_2) \mid \lambda_1<\lambda_{1}^{SR}(\lambda_{2}),\lambda_2<\lambda_{2}^{SR}(\lambda_{1}) \rbrace .
\end{equation}
With respect to (\ref{19}), for each $i=1,2$, in order to find $\lambda_{i}^{SR}$, $\sum\limits_{h=1}^m {\boldsymbol{\alpha}(h)p_{i}^{suc}(h)}$ should be calculated when node $N_i$ is saturated. In this case that we have $K=0$, the backoff stage vector is always equal to zero. So, in each $MC_i$ the number of phases is $m_0=m=2$. It means that each level $\ell$ has two phases $(\boldsymbol{0},0)$ and $(\boldsymbol{0},1)$, which correspond to $h=1,2$, respectively. In other words, the second phase of each level ($h=2$) indicates that the other node (i.e., $N_j$), has at least one packet but the first phase represents the situation that it is empty. So, for these phases, $p_i^{suc}(h)$ which denotes the probability of a successful transmission for node $N_i$, can be obtained as
\vspace{-0.3cm}
\begin{equation}\label{22}
p_i^{suc}(h)=p_i (1-p_j I(h=2)).
\end{equation}

In the next step, we should calculate matrix $A$ and its stationary probability vector $\boldsymbol{\alpha}$, when node $N_i$ is saturated. As discussed before, elements of $A$ are related to $z_j$. So, first we must find $z_j$ by assuming that node $N_i$ is saturated ($z_i=0$). As mentioned in proof of Theorem 3, by this assumption states of ${MC}_j$ ($j \neq i$) in which $\hat{q}_i=0$ become transient. So, ${MC}_j$ becomes a one-dimensional Markov chain. By obtaining the stationary distribution of the queue lengths of node $N_j$ ($\pi_j(.)$), $z_j$ is derived as
\begin{equation}\label{24}
z_j=\dfrac {\pi_j(1)} {1-\pi_j(0)}=\dfrac {p_j (1-p_i)-\lambda_j} {(1-\lambda_j)p_j(1-p_i)}.
\end{equation}
So, from (\ref{16}), matrix $A$ and its stationary probability vector are derived as
\begin{equation}\label{26}
A =
 \setlength{\arraycolsep}{9pt}
\left(
	\begin{array}{ll} 
		1-\lambda_j  & \lambda_j  \\
		p_j (1-p_i)-\lambda_j & 1-p_j (1-p_i)+\lambda_j
	\end{array}
\right),
\end{equation}
\begin{equation}\label{27}
\boldsymbol{\alpha}=
 \setlength{\arraycolsep}{9pt}
\left(
	\begin{array}{ll} 
 1- \dfrac{\lambda_j} {p_j (1-p_i)} & \dfrac{\lambda_j} {p_j (1-p_i)}
	\end{array}
\right).
\end{equation}
Then, by using (\ref{22}) and (\ref{27}), $\lambda_{i}^{SR}$ can be obtained as
\ifCLASSOPTIONdraftclsnofoot
\begin{equation}\label{28}
\lambda_{i}^{SR}=(1- \dfrac{\lambda_j} {p_j (1-p_i)})p_i+\dfrac{\lambda_j} {p_j (1-p_i)} p_i (1-p_j)=
p_i (1-\dfrac{\lambda_j} {1-p_i}).
\end{equation}
\else
\begin{multline}\label{28}
\lambda_{i}^{SR}=(1- \dfrac{\lambda_j} {p_j (1-p_i)})p_i+\dfrac{\lambda_j} {p_j (1-p_i)} p_i (1-p_j)= \\
p_i (1-\dfrac{\lambda_j} {1-p_i}).
\end{multline}
\fi
So, the stability region is derived as
\ifCLASSOPTIONdraftclsnofoot
\begin{equation}\label{29}
SR=\lbrace (\lambda_1,\lambda_2) \mid \lambda_1<p_1 (1-\dfrac{\lambda_2} {1-p_1}),\lambda_2<p_2 (1-\dfrac{\lambda_1} {1-p_2}) \rbrace ,
\end{equation}
\else
\small
\begin{equation}\label{29}
SR=\lbrace (\lambda_1,\lambda_2) \mid \lambda_1<p_1 (1-\dfrac{\lambda_2} {1-p_1}),\lambda_2<p_2 (1-\dfrac{\lambda_1} {1-p_2}) \rbrace ,
\end{equation}
\normalsize
\fi
which equals the exact stability region derived in \cite{R1}.

\section{Stability Region of the Network with D-MAP}

As mentioned before, in our network for each $i \in \lbrace 1,\allowbreak \dots,\allowbreak n \rbrace$ the arriving process of node $N_i$ is a D-MAP with a Markov chain with $c_i$ states. For this Markov chain, matrices $D_{0,i}$ and $D_{1,i}$ represent the transition probabilities between different states without and with an arrival, respectively. Moreover, we defined $\lambda_{i}(u)=\sum_{v}{d_{1,i}(u,v)}$ as the arrival probability for node $N_i$ when it is in state $u$, and $\lambda_i=\sum_{u}{\pi^{A}_i(u) \lambda_{i}(u)}$ as the average arrival rate of packets at node $N_i$. Notation $\pi^{A}_i(u)$ denotes the stationary probability of being in state $u$ of the arrival Markov chain corresponding to node $N_i$.

For modeling the slotted Aloha network with D-MAPs similar to the previous sections, the set of states should be slightly modified. It means that we model each node $N_i$ by a Markov chain ${MC}_i^{D}$ that has a set of states different from the set defined in Section III. Let us define $S_{i}^{D}=\lbrace (q_i,(\boldsymbol{b}_i,\hat{\boldsymbol{q}}_{-i},u_i)) \rbrace$ as the set of states for ${MC}_i^{D}$, where $1 \leq u_i \leq c_i$ indicates that at this time slot, the arrival Markov chain corresponding to node $N_i$ is at the $u_i^{th}$ state. It is worth noting that in order to reduce the number of states, in each ${MC}_i^{D}$ we only model the arrival process of node $N_i$ and assume approximately that all other nodes ($N_j$s for $j \neq i$) have Bernoulli arrival processes with rate $\lambda_j$.

It is clear that D-MAPs are single-arrival processes. It means that at each time slot the queue length of node $N_i$ (i.e., $q_i$) can increase at most one unit. So, similar to the case that nodes have Bernoulli arrival processes, the obtained Markov chain (${MC}_i^{D}$) is a homogenous QBD with transition probability matrix ($P^{D}$) of the form (\ref{2}). The number of phases in level $0$ and level $\ell$ ($\ell \geq 1$) are $m_{0}^{D}={(K+2)}^{n-1} c_i$ and $m^{D} = m_{\ell}^{D}=(K+1){(K+2)}^{n-1} c_i$, respectively. We line up phases of each level in lexicographic order and consider $\varphi_{i}^{D} (\ell,h),h=1,\ldots,m_{\ell}^{D}$ as the $h^{th}$ phase of the $\ell^{th}$ level of ${MC}_i^{D}$. 

Transition probabilities and inner submatrices of $P^{D}$ can be found by a procedure similar to one used in Section III. Let us denote the inner submatrices of $P^{D}$ by $A_k^{D}$ and $B_k^{D}$ for $k \in \lbrace 0,1,2 \rbrace$. We also define $A^{D}$ as the stochastic matrix $A^{D}=A_0^{D}+A_1^{D}+A_2^{D}$. Then, we take the steps analog to ones in Section IV to find the stability region. Here, it is important to be clear on what we mean by the term 'stability region' in a network with D-MAPs. In such networks, we consider stability region as the set of all irreducible D-MAPs that keep the network stable. It means that we are interested in determining the set of matrices $\lbrace D_{0,1}$, $D_{1,1},\ldots, D_{0,n}$, $D_{1,n} \rbrace$ by which for all nodes the condition of balanced input-output rates is satisfied and the arrival Markov chains of all D-MAPs become irreducible.

It is easy to show that in this model, similar to previous one, the stability condition is equivalent to positive recurrency of QBDs. In other words, for a given set of matrices $\lbrace D_{0,1},\allowbreak D_{1,1},\allowbreak \ldots,\allowbreak D_{0,n},\allowbreak D_{1,n} \rbrace$ that makes the arrival Markov chains irreducible, if and only if all QBDs are positive recurrent, we can conclude that this set of arrival processes is in the stability region. So, in order to determine the stability region, Theorem 1 will be the main tool. In our model, each ${MC}_i^{D}$ and its corresponding stochastic matrix $A^{D}$ are irreducible. It is due to the fact that arrival Markov chains corresponding to D-MAPs are considered to be irreducible. So, it can be derived from Theorem 1 that the stability region comprises all sets of matrices for which the condition $\mu<0$ is satisfied for all QBDs (i.e., ${MC}_1^{D},\ldots,{MC}_n^{D}$). Thus, in order to find the stability region, first we should calculate the value of $\mu=\boldsymbol{\alpha}^{D}(A_0^{D}-A_2^{D})\boldsymbol{1}$ corresponding to each ${MC}_i^{D}$ in terms of transition probability matrices of arrival processes. $\boldsymbol{\alpha}^{D}$ is the stationary probability vector of $A^{D}$. $A^{D}$ is a square matrix of order $m^{D}$ that its elements can be calculated in the same way as (\ref{16}). Moreover, in order to calculate $(A_0-A_2)\boldsymbol{1}$ in Section IV, Theorem 2 was used. But here, we use the following theorem:

\begin{theorem}
In each ${MC}_{i}^{D}$, $(A_0^{D}-A_2^{D})\boldsymbol{1}$ is a column vector of order $m^{D}$ that its $h^{th}$ element is equal to $\lambda_{i}(h)-p_{i}^{suc,D}(h)$, where $\lambda_{i}(h)=\lambda_{i,u_i}$ is the arrival probability of node $N_i$ when the current state is $s=(\ell,\varphi_i^{D}(\ell,h))$ and the arrival Markov chain corresponding to $N_i$ is at $u^{th}_i$ state. Moreover, $p_{i}^{suc,D}(h)$ is the probability that $i^{th}$ node has a successful transmission when the current state is $s^{D}=(\ell,\varphi_i^{D}(\ell,h)), \ell \geq 2$.
\end{theorem}

\begin{IEEEproof}
The proof of this theorem is very similar to Theorem 2, hence omitted. 
\end{IEEEproof}

According to Theorem 4, $\mu$ which is defined as $\mu=\boldsymbol{\alpha}^{D}(A_0^{D}-A_2^{D})\boldsymbol{1}$ for ${MC}_{i}^{D}$, is obtained as follows:
\begin{equation}\label{30}
\mu=\sum\limits_{h=1}^{m^{D}} {\boldsymbol{\alpha}^{D}(h)\lambda_{i}(h)}-\sum\limits_{h=1}^{m^{D}} {\boldsymbol{\alpha}^{D}(h)p_{i}^{suc,D}(h)},
\end{equation}
where $\boldsymbol{\alpha}^{D}(h)$ is the stationary probability of being in $h^{th}$ phase of a Markov chain with $A^{D}$ as its transition probability matrix. Phases with the same $u_i$ have equal $\lambda_{i}(h)$, i.e., $\lambda_{i,u_i}$. The arrival Markov chain of node $N_i$ which determines $u_i$ is not affected by any other things and it works independently. So, 
 \begin{equation}\label{31}
\sum_{h|u_i} {\boldsymbol{\alpha}^{D}(h)}=\pi^{A}_i(u_i),
\end{equation}
where $h|u_i$ means that the summation is over all phases in which the arrival Markov chain corresponding to node $N_i$ is at $u_i^{th}$ state. As a result we have 
\ifCLASSOPTIONdraftclsnofoot
\begin{equation}\label{32}
\mu=\sum\limits_{u_i=1}^{c_i} {\pi^{A}_i(u_i)\lambda_{i,u_i}}-\sum\limits_{h=1}^{m^{D}} {\boldsymbol{\alpha}^{D}(h)p_{i}^{suc,D}(h)}=
\lambda_i-\sum\limits_{h=1}^{m^{D}} {\boldsymbol{\alpha}^{D}(h)p_{i}^{suc,D}(h)}.
\end{equation}
\else
\begin{multline}\label{32}
\mu=\sum\limits_{u_i=1}^{c_i} {\pi^{A}_i(u_i)\lambda_{i,u_i}}-\sum\limits_{h=1}^{m^{D}} {\boldsymbol{\alpha}^{D}(h)p_{i}^{suc,D}(h)}=\\
\lambda_i-\sum\limits_{h=1}^{m^{D}} {\boldsymbol{\alpha}^{D}(h)p_{i}^{suc,D}(h)}.
\end{multline}
\fi

\begin{theorem}
When node $N_i$ is saturated, if we consider a similar network that nodes have Bernoulli arrival processes with equivalent average arrival rates ($\boldsymbol{\lambda}$) and model it with ${MC}_1,\allowbreak \ldots,\allowbreak{MC}_n$ as described in Section III, we have
\begin{equation}\label{33}
\sum\limits_{h=1}^{m^{D}} {\boldsymbol{\alpha}^{D}(h)p_{i}^{suc,D}(h)}=\lambda_{i}^{SR}({\boldsymbol{\lambda}}_{-i}),
\end{equation}
where $\lambda_{i}^{SR}({\boldsymbol{\lambda}}_{-i})$ is the value of $\lambda_i$ which makes ${MC}_i$ be in the border of instability.
\end{theorem}

\begin{IEEEproof}
See Appendix.
\end{IEEEproof} 

In the following, we call the network in which the nodes have Bernoulli arrival processes with equivalent average arrival rates, as `rate-equivalent network'. It can be concluded from (\ref{32}), Theorem 1, and Theorem 5, that the necessary and sufficient condition for network to be stable is $\lambda_i<\lambda_{i}^{SR}({\boldsymbol{\lambda}}_{-i})$ for all $i \in \lbrace 1,\ldots,n \rbrace$. So, in determining the stability region with our proposed method only the average arrival rates are effective and the statistics of arrival processes do not need to be considered.

\section{Numerical Results}
In this section, we present the numerical results of our proposed analytical approach in different conditions. In order to show the accuracy of our analysis, we compare our results with simulation ones. Our simulation is done in MATLAB environment. In simulations we have considered a large time interval and compute the ratio of the number of successfully transmitted packets of all nodes on the number of newly arrived packets at the same time interval. In stable conditions the ratio equals one, but by increasing the packet arrival rates, the ratio becomes smaller than one indicating that the network becomes unstable. So, in order to find the boundary of the stability region we fix the arrival rates of $n-1$ nodes and increase the arrival rate of the last node until the network becomes unstable. In Fig. \ref{backoff45}, we have focused on a two-node slotted Aloha network with Bernoulli arrival processes and illustrated its stability region when binary exponential backoff mechanism ($r=2$) with different cutoff stages (i.e., $K=0,1,2,3$) is used. In Fig. \ref{backoff4}, a case with low transmission probabilities, i.e., $p_1=p_2=0.2$ and in Fig. \ref{backoff5}, another case with high transmission probabilities, i.e., $p_1=p_2=0.8$, have been considered. It is observed that there is a good match between analytical results and simulation ones. As mentioned in Section IV, when the network consists of two nodes and there is no exponential backoff mechanism (i.e., $n=2, K=0$) our analytical results are exactly the same as the explicit form of the stability region of this network, presented in \cite{R1}.

\ifCLASSOPTIONdraftclsnofoot
\begin{figure*}
\centering
\subfloat[\vspace{-0.15cm} $p_1=p_2=0.2$.]{\includegraphics[width=3in]{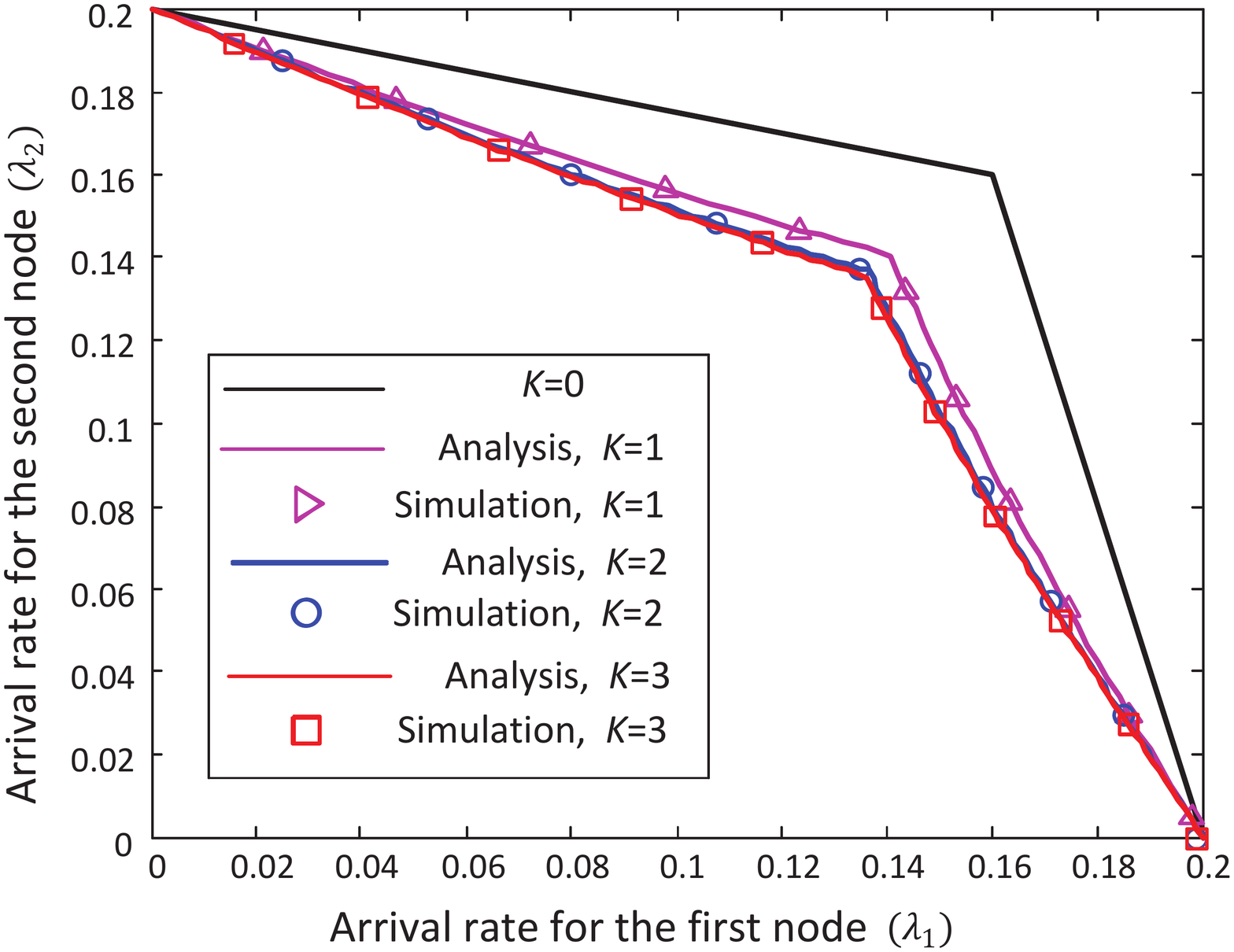}
\label{backoff4}}
\subfloat[\vspace{-0.15cm} $p_1=p_2=0.8$.]{\includegraphics[width=3in]{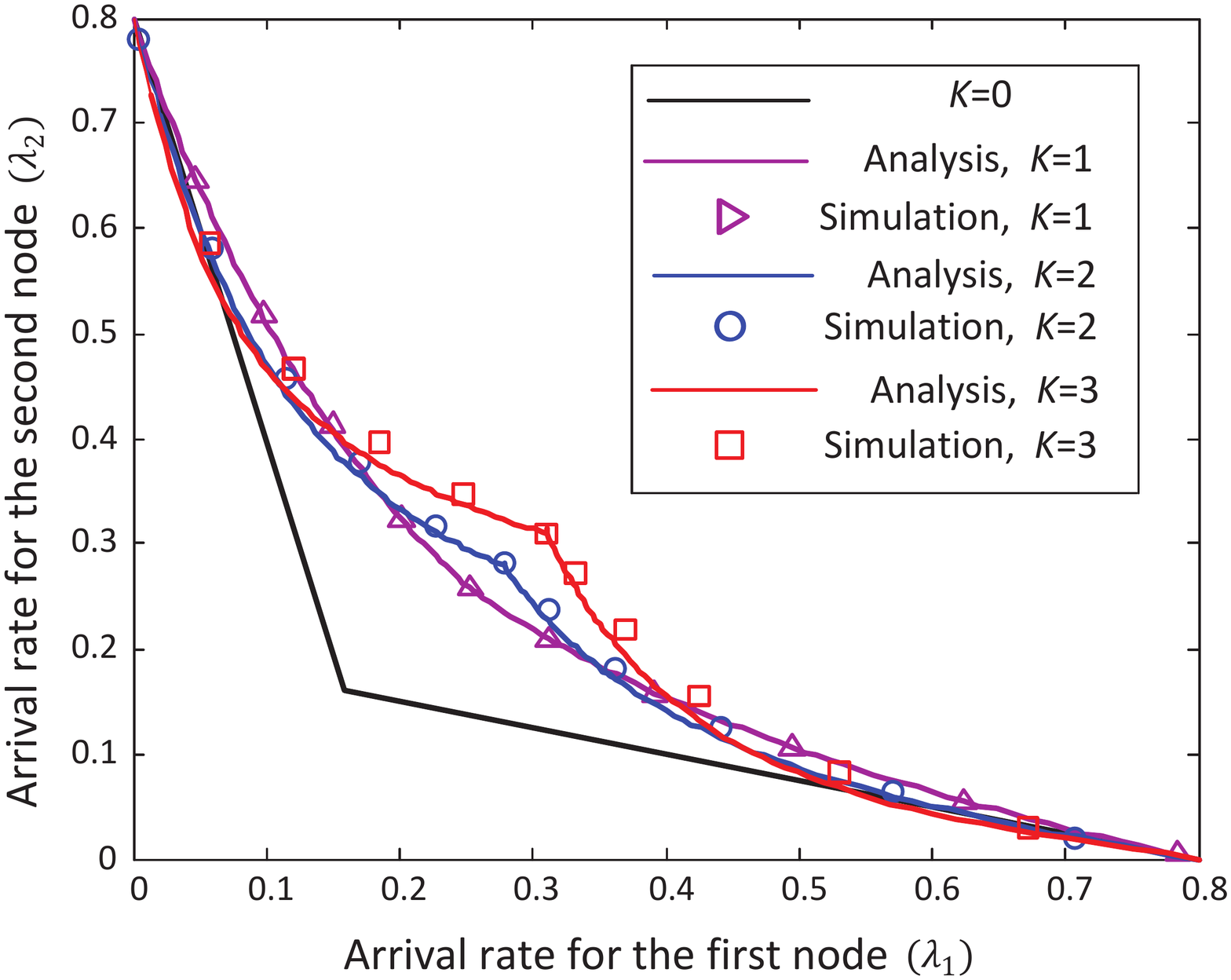}
\label{backoff5}}
\caption{Stability region of a two-node network for different cutoff stages, $r=2$.}
\vspace{-0.7cm}
\label{backoff45}
\end{figure*}
\else
\begin{figure}
\centering
\subfloat[\vspace{-0.15cm} $p_1=p_2=0.2$.]{\includegraphics[width=0.9\columnwidth,height=2in]{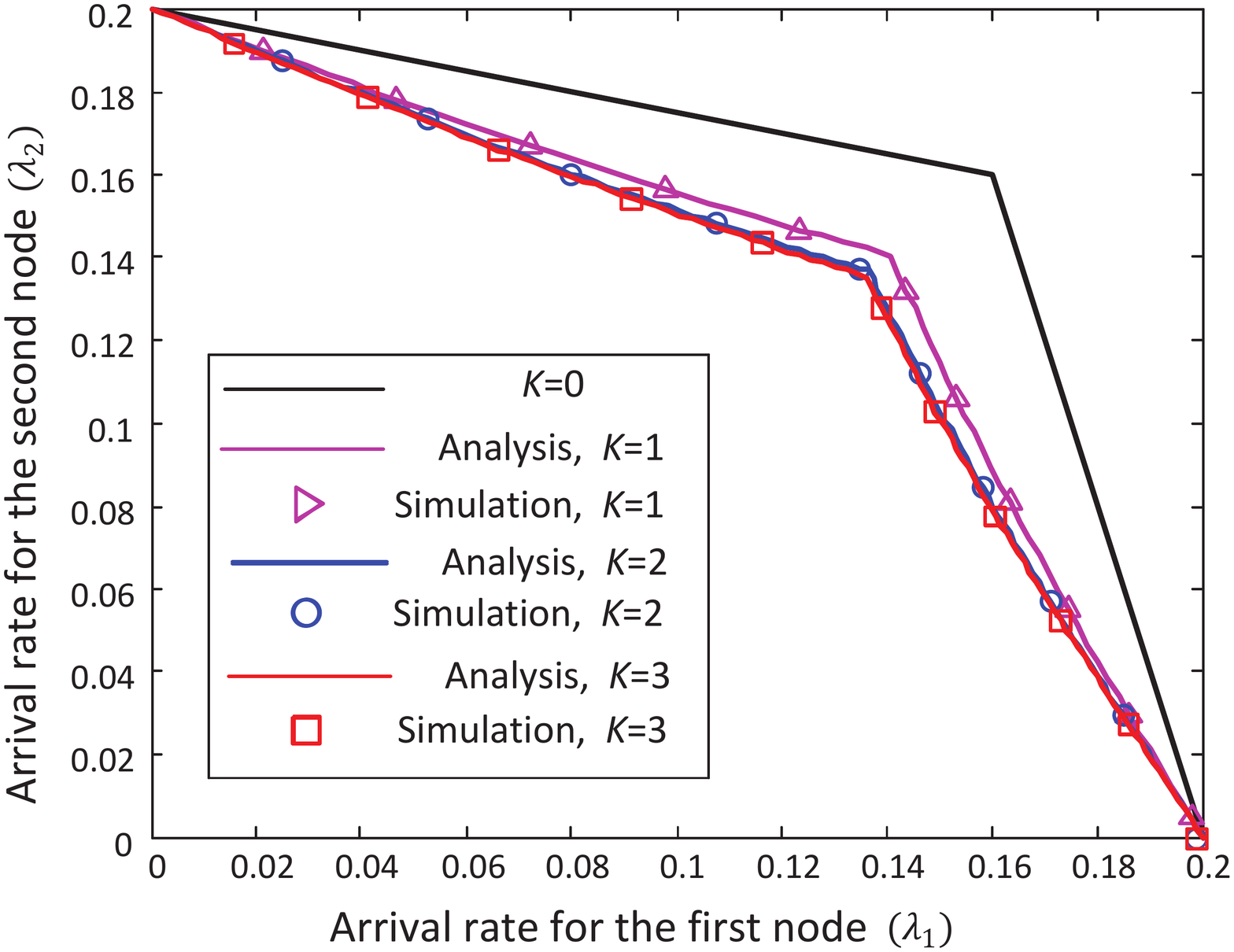}
\label{backoff4}}
\hfil
\subfloat[\vspace{-0.15cm} $p_1=p_2=0.8$.]{\includegraphics[width=0.9\columnwidth,height=2in]{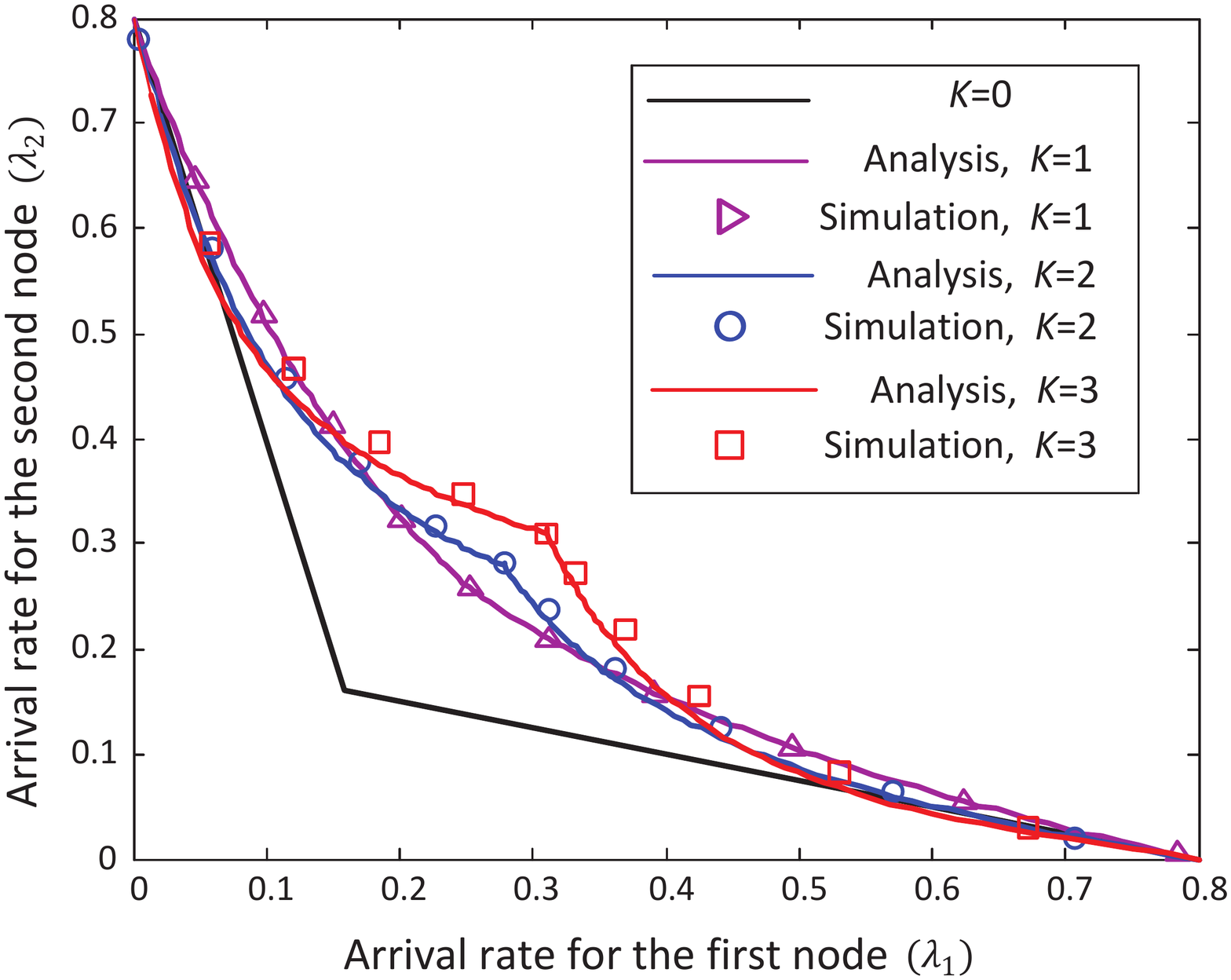}
\label{backoff5}}
\vspace{0.05cm}
\caption{Stability region of a two-node network for different cutoff stages, $r=2$.}
\vspace{-0.5cm}
\label{backoff45}
\end{figure}
\fi

%
%

Moreover, we can observe that when transmission probabilities of both nodes are low (see Fig. \ref{backoff4}), as a result of low probability of collision, using binary exponential backoff mechanism is not a useful method for expanding the stability region. While, Fig. \ref{backoff5} shows that in the case with high transmission probabilities, using backoff mechanism improves the stability region in high collision points significantly.

\begin{figure}
\centering

\ifCLASSOPTIONdraftclsnofoot
\includegraphics[width=3in]{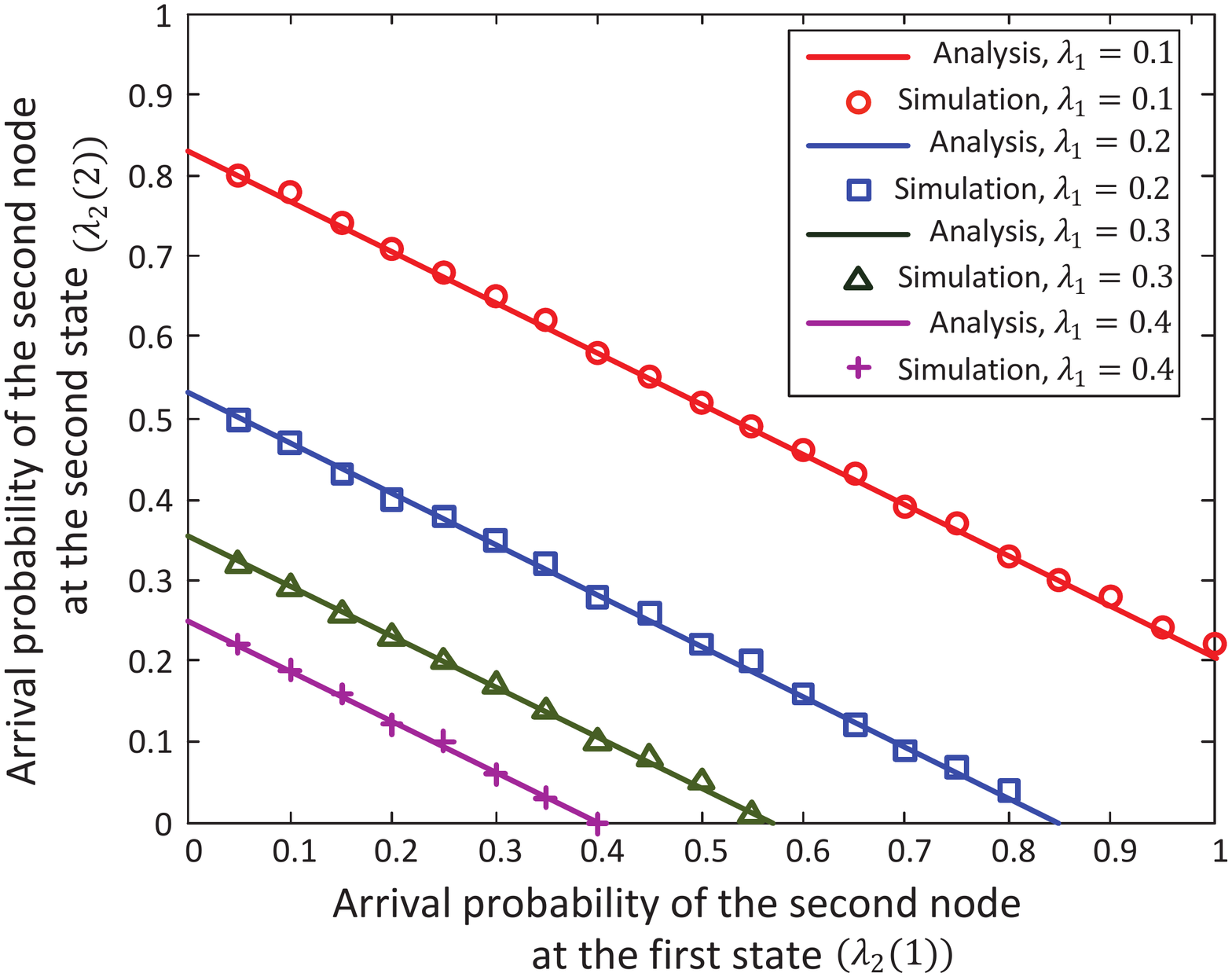}
\vspace{-0.8cm}
\else
\includegraphics[width=0.9\columnwidth,height=2in]{Fig10.pdf}
\vspace{-0.3cm}
\fi

\caption{Arrival probabilities that make the network with D-MAPs in the border of instability, $p_1=p_2=0.8$, $r=2$, $K=1$.}
\ifCLASSOPTIONdraftclsnofoot
\vspace{-0.8cm}
\else
\vspace{-0.3cm}
\fi
\label{backoff10}
\end{figure}

As discussed in Section V, our modeling approach shows that in determining the stability region of networks with D-MAPs, only the average arrival rates of nodes are operative and the statistics of arrival processes do not need to be considered. In order to show the accuracy of this significant result, we consider a two-node network with D-MAPs that uses slotted Aloha protocol with one-stage binary exponential backoff mechanism ($r=2,K=1$) and symmetric initial transmission probabilities $p_1=p_2=0.8$. We assume that for $i=1,2$, the D-MAP corresponding to node $N_i$ has two states and is specified by following matrices $D_{0,i}$ and $D_{1,i}$:
\ifCLASSOPTIONdraftclsnofoot
\vspace{-0.2cm}
\begin{equation} \label{43}
D_{0,i}=
 \setlength{\arraycolsep}{9pt}
\left(
	\begin{array}{ll} 
	0.2(1-\lambda_{i}(1))  & 0.8(1-\lambda_{i}(1))  \\
		0.5(1-\lambda_{i}(2)) & 0.5(1-\lambda_{i}(2))
	\end{array}
\right),
D_{1,i}=
 \setlength{\arraycolsep}{9pt}
\left(
	\begin{array}{ll} 
	0.2\lambda_{i}(1) & 0.8\lambda_{i}(1)  \\
		0.5\lambda_{i}(2) & 0.5\lambda_{i}(2)
	\end{array}
\right),
\end{equation}
\else
\begin{align} \label{43}
D_{0,i}&=
 \setlength{\arraycolsep}{9pt}
\left(
	\begin{array}{ll} 
	0.2(1-\lambda_{i}(1))  & 0.8(1-\lambda_{i}(1))  \\
		0.5(1-\lambda_{i}(2)) & 0.5(1-\lambda_{i}(2))
	\end{array}
\right),\nonumber\\
D_{1,i}&=
 \setlength{\arraycolsep}{9pt}
\left(
	\begin{array}{ll} 
	0.2\lambda_{i}(1) & 0.8\lambda_{i}(1)  \\
		0.5\lambda_{i}(2) & 0.5\lambda_{i}(2)
	\end{array}
\right),
\end{align}
\fi
where $\lambda_{i}(u)$ denotes the arrival probability for node $N_i$ when it is in state $u$. It is clear that irrespective of the values of $\lambda_{i}(u)$s, the stationary distribution of arrival Markov chains corresponding to D-MAPs are $\boldsymbol{\pi}^{A}_i=(0.385, 0.615)$. So, the average arrival rate of node $N_i$ is $\lambda_i=0.385 \lambda_{i}(1) + 0.615 \lambda_{i}(2)$. In Fig. \ref{backoff10} we have considered four different pairs of arrival probabilities for node $N_1$, i.e., ($\lambda_{1}(1),\lambda_{1}(2)$), which lead to average arrival rates $\lambda_1=0.1,0.2,0.3,0.4$, respectively. For each of these cases, we keep the D-MAP of node $N_1$ constant and determine by simulation, the pairs of ($\lambda_{2}(1),\lambda_{2}(2)$) which make network in the border of instability. Fig. \ref{backoff5} shows that in rate-equivalent network with Bernoulli arrival processes, for $\lambda_1=0.1,0.2,0.3,0.4$, the average arrival rates of node $N_2$ that make the network in the border of instability are $\lambda_2^{max}=0.51,0.326,0.219,0.154$, respectively. In Fig. \ref{backoff10}, the pairs of arrival probabilities of node $N_2$ that lead to average arrival rates equal to $\lambda_2^{max}$ are shown by line and the pairs of arrival probabilities that make the network in the border of instability are shown by markers. So, the results confirm that only the average arrival rates play the key role in the network.

Now, in order to show the effects of exponential backoff mechanism on stability region of the network, we introduce two metrics. The first metric is the $n$-dimensional volume of the stability region which is equivalent to area of the stability region for $n=2$. In  Fig. \ref{backoff78}, we have focused on networks with two and three nodes ($n=2,3$), respectively, in symmetric cases and shown the $n$-dimensional volume of the stability region ($V$) versus backoff factor ($r \geq 1$) for different transmission probabilities ($p_i=p, \forall i \in \lbrace 1,\ldots,n \rbrace$), when the cutoff stage is $K=1$. It can be observed that for high transmission probabilities by increasing the value of backoff factor up to a certain optimum ($r_{opt}^{v}$), the volume of stability region increases and then it starts to decrease.

In symmetric two-node networks, the two-dimensional volume (i.e., area) of stability region for slotted Aloha protocol without backoff, which is given by $V=p^{2}(1-p)$, has its maximum value of $V=0.148$ at $p=2/3$. While, Fig. \ref{backoff7}, shows that for such a protocol with exponential backoff mechanism and $K=1$, the maximum area of stability region is $V=0.213$ and it occurs at $p=1$ with $r=2.6$. It means that the exponential backoff mechanism is able to improve the maximum area of the stability region of a two-node network more than 40\%.

In Fig. \ref{backoff8}, we show that in a three-node network the optimum values of backoff factor for the cases $p=1/5,\allowbreak1/3,\allowbreak 1/2,\allowbreak 4/5,\allowbreak 1$, are equal to $r_{opt}^{v}=1,\allowbreak 1,\allowbreak 1,\allowbreak 1.5,\allowbreak 4.5$, respectively. This figure shows that the exponential backoff mechanism with $K=1$ improves the maximum volume of the stability region of a three-node network almost 22\%.

\ifCLASSOPTIONdraftclsnofoot
\begin{figure*}
\centering
\subfloat[\vspace{-0.15cm} $n=2$.]{\includegraphics[width=3in]{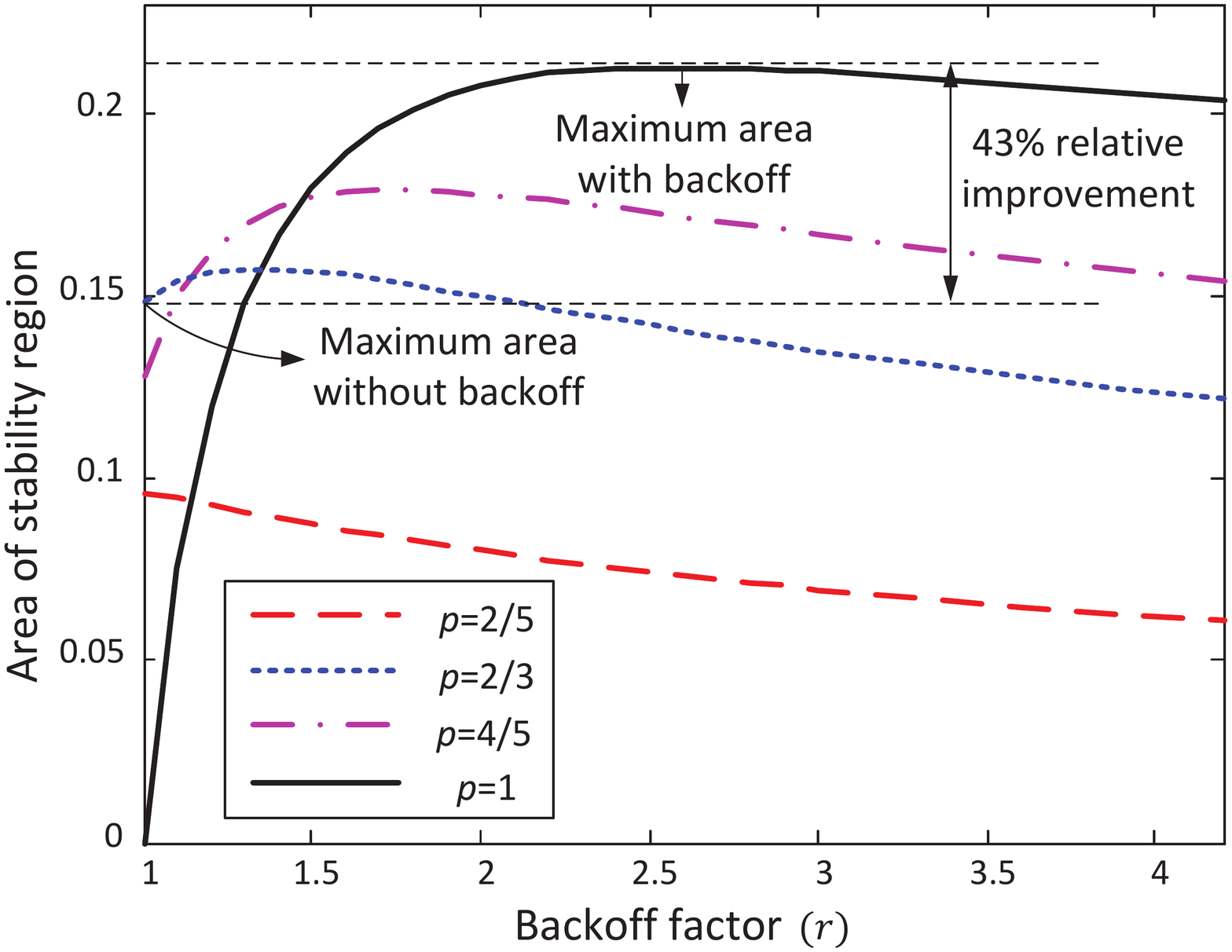}
\label{backoff7}}
\subfloat[\vspace{-0.15cm} $n=3$.]{\includegraphics[width=3in]{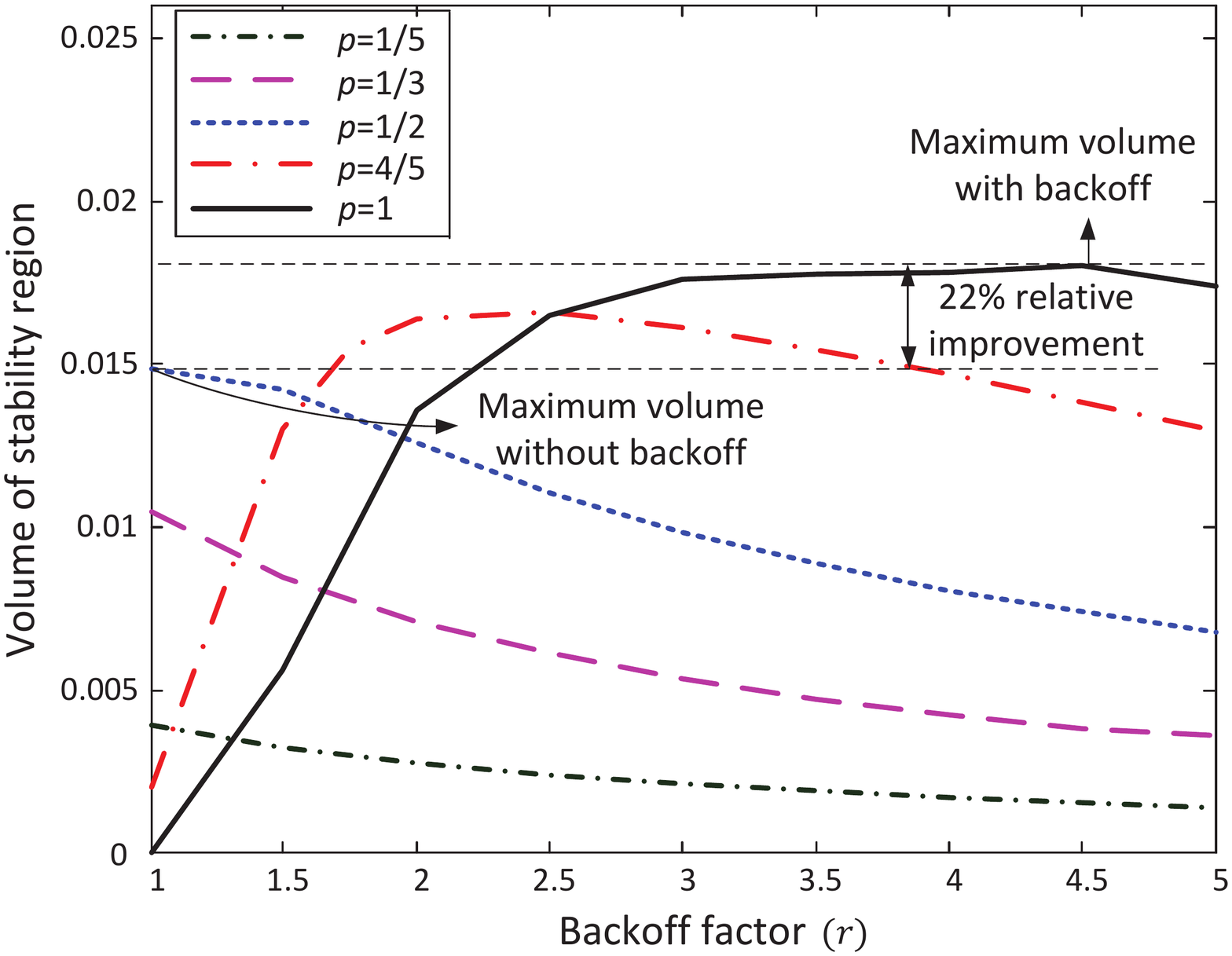}
\label{backoff8}}
\caption{$n$-dimensional volume of stability region vs. backoff factor, $K=1$.}
\vspace{-0.7cm}
\label{backoff78}
\end{figure*}
\else
\begin{figure}
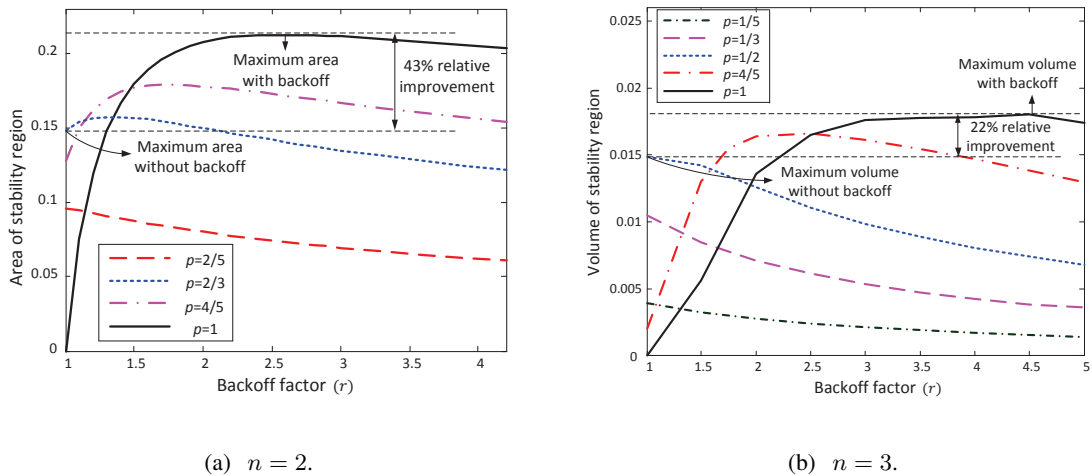

\centering
\subfloat[\vspace{-0.15cm} $n=2$.]{\includegraphics[width=0.9\columnwidth,height=2in]{Fig6.pdf}
\label{backoff7}}
\hfil
\subfloat[\vspace{-0.15cm} $n=3$.]{\includegraphics[width=0.9\columnwidth,height=2in]{Fig7.pdf}
\label{backoff8}}
\vspace{0.05cm}
\caption{$n$-dimensional volume of stability region vs. backoff factor, $K=1$.}
\vspace{-0.3cm}
\label{backoff78}
\end{figure}
\fi

The second metric that we use to evaluate the performance of the exponential backoff mechanism is \emph{sum saturation throughput}, which is the summation of the throughputs achieved by nodes, when all nodes are saturated. In Fig. \ref{backoff9} we have considered a symmetric network with four nodes (i.e., $n=4$) and shown the sum saturation throughput versus backoff factor for different transmission probabilities ($p$), when the cutoff stage is $K=1$. It is easy to show that in symmetric four-node networks, the sum saturation throughput for slotted Aloha protocol without backoff is given by $4p{(1-p)}^{3}$ and has its maximum value of $0.4219$ at $p=1/4$. While, Fig. \ref{backoff9} shows that for such a protocol with exponential backoff mechanism and $K=1$, the maximum sum saturation throughput is $0.571$ and it occurs at $p=1$. It means that the exponential backoff mechanism with $K=1$ is able to improve the maximum sum saturation throughput of a four-node network about 35\%. In Fig. \ref{backoff11}, we have shown the maximum sum saturation throughput of symmetric networks versus number of nodes for different cutoff stages. This figure shows that by using exponential backoff mechanism with $K=2$, we can improve the maximum sum saturation throughput of symmetric networks about 100\% and come nearly close to having one successful transmission at each time slot, which is the optimal throughput obtained by centralized scheduling.

%
%
%
%
%
%

\ifCLASSOPTIONdraftclsnofoot
\begin{figure*}
\centering
\subfloat[\vspace{-0.15cm}]{\includegraphics[width=3in,height=2.2in]{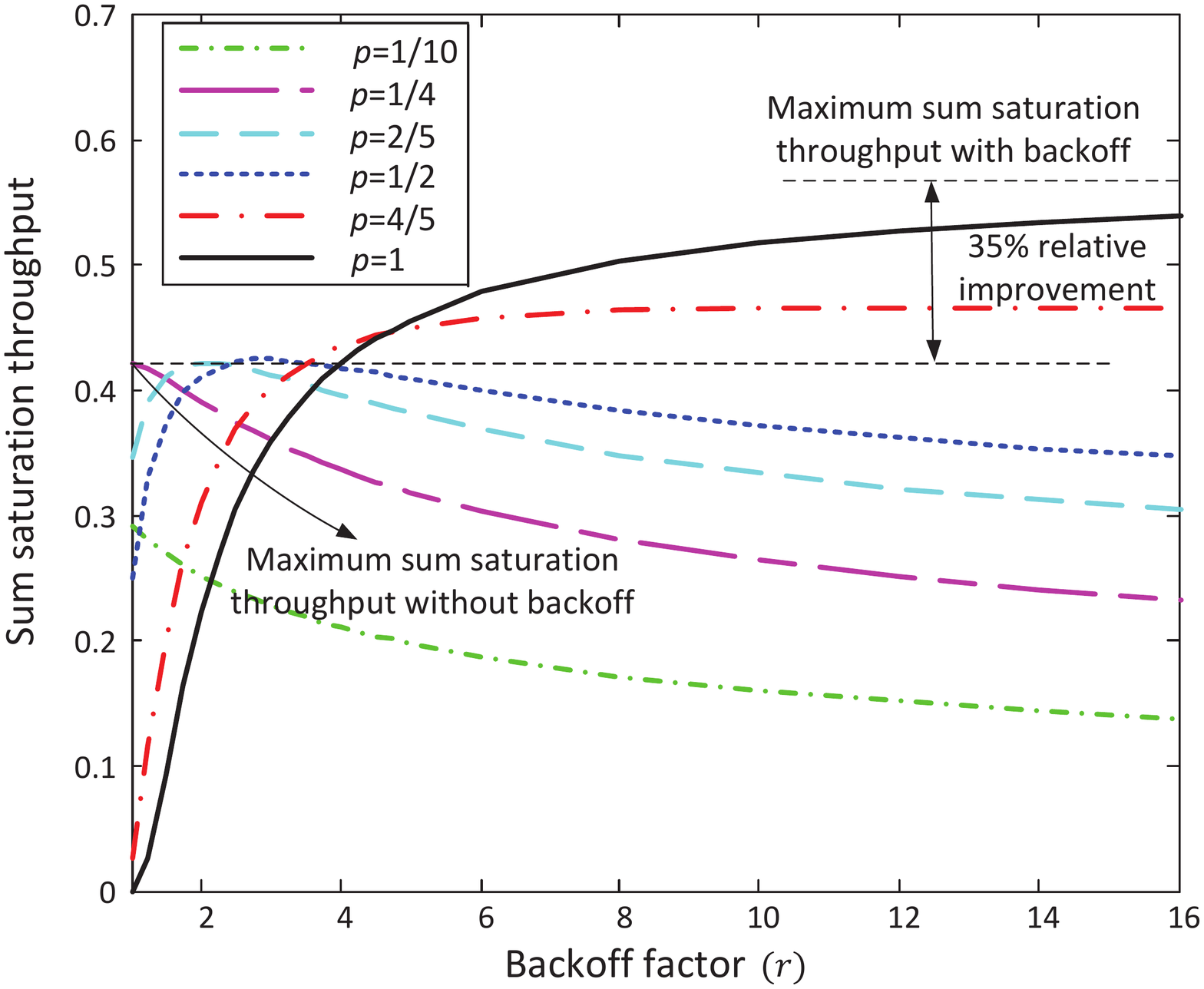}
\label{backoff9}}
\subfloat[\vspace{-0.15cm}]{\includegraphics[width=3in,,height=2.2in]{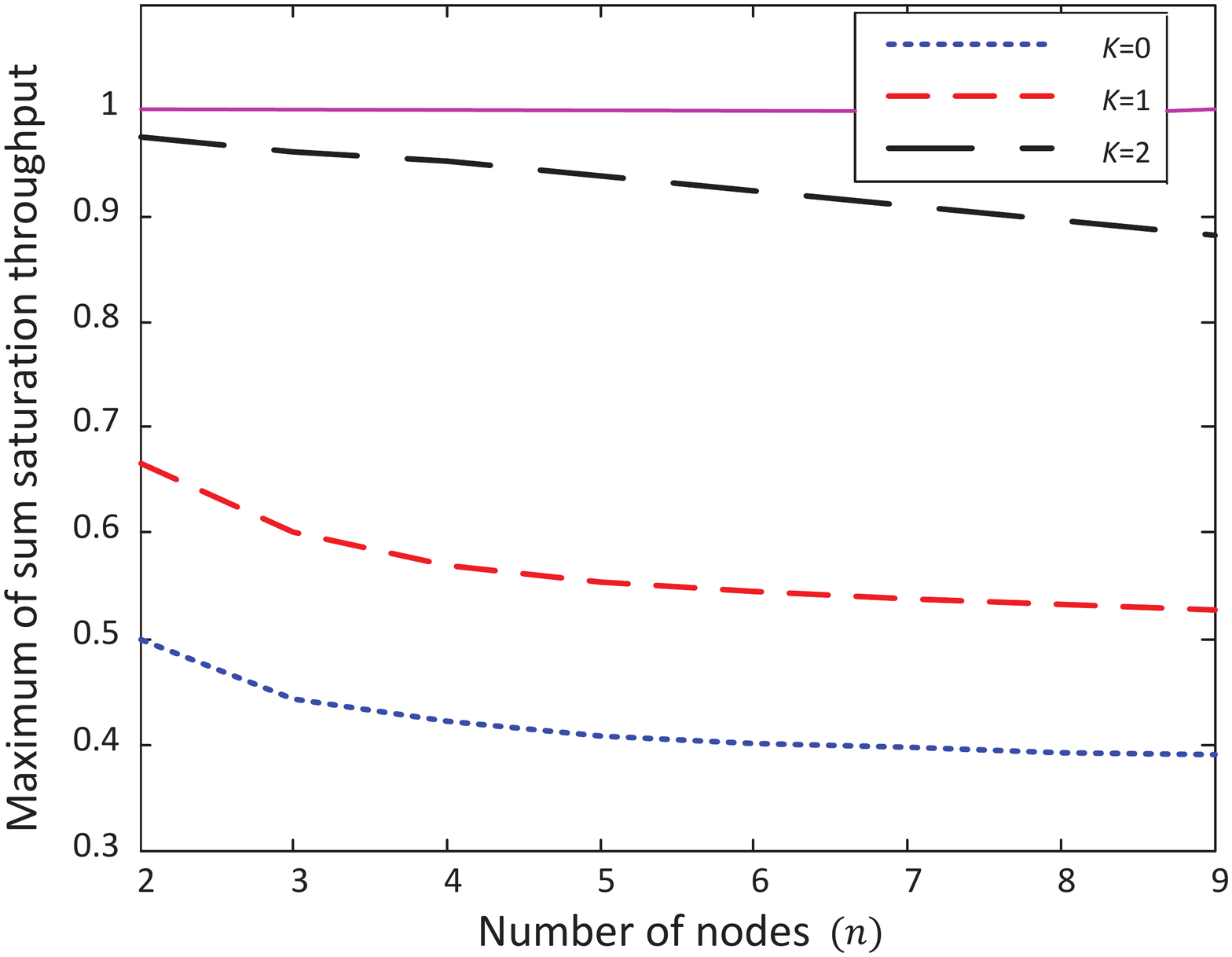}
\label{backoff11}}
\caption{(a) Sum saturation throughput, $n=4,K=1$. (b) Maximum of sum saturation throughput vs. number of nodes.}
\vspace{-0.7cm}
\label{backoff911}
\end{figure*}
\else
\begin{figure}
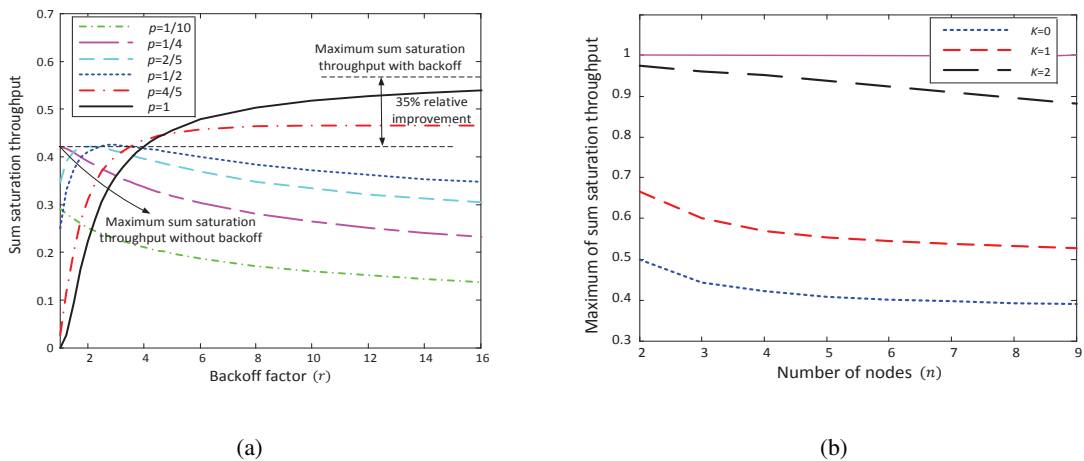

\centering
\subfloat[\vspace{-0.15cm}]{\includegraphics[width=0.9\columnwidth,height=2in]{Fig9.pdf}
\label{backoff9}}
\hfil
\subfloat[\vspace{-0.15cm}]{\includegraphics[width=0.9\columnwidth,height=2in]{Fig11.pdf}
\label{backoff11}}
\vspace{0.05cm}
\caption{(a) Sum saturation throughput, $n=4,K=1$. (b) Maximum of sum saturation throughput vs. number of nodes.}
\vspace{-0.5cm}
\label{backoff911}
\end{figure}
\fi

\section{Conclusion}
Stability analysis of a buffered slotted Aloha-based network with $K$-exponential backoff mechanism and D-MAPs was studied in this paper. First, we focused on Bernoulli arrival process as a simple example of this nearly general family of arrival processes and presented an approximate analytical model. In this model, we mapped the states of each node on a Markov chain. Moreover, in order to track the interaction among queues and the memory of system which is caused by interaction among queues as well as exponential backoff mechanism, we considered an indication of the status of the other nodes as well as backoff stage vector in each Markov chain. Second, based on the fact that each Markov chain is a QBD process and with the help of theorems on stability of QBDs, we found an approximation for the stability region of the slotted Aloha-based network with $K$-exponential backoff mechanism and Bernoulli arrival processes. Then we generalized our model for networks in which nodes have D-MAPs and showed that in determining the stability region with our proposed method the statistics of the arrival processes do not need to be considered. Finally we showed that our obtained analytical results are highly matched with simulation ones. We also introduced two new metrics to evaluate the performance of exponential backoff mechanism. Our results emphasize that by proper setting of the backoff factor and cutoff stage, the stability region of the network can be enhanced significantly compared to the network without exponential backoff and sum saturation throughput of the network may approach to centralized scheduling.

\appendix[Proof of Theorem 5]
Before we prove this theorem, let us prove a useful lemma.
\newtheorem {lemma}{Lemma}
\begin{lemma}
When node $N_i$ is saturated, if we model both real network and rate-equivalent network with QBDs and consider $A$ and $A^D$ as the stochastic matrices corresponding to ${MC}_i$ in rate-equivalent network and ${MC}_i^{D}$ in real network, respectively, we have
\begin{equation}\label{34}
A^D=A \otimes D_i,
\end{equation}
where $D_i=D_{0,i}+D_{1,i}$ and the symbol $\otimes$ represents the Kronecker product.
\end{lemma}

\begin{IEEEproof}
When we consider $N_i$ to be saturated, analysis of ${MC}_j^{D}$s and as a result $z_j$s becomes independent of arrival process of node $N_i$. So, the elements of $A^{D}$ become independent of whether a packet is arriving at node $N_i$ or not. In this situation, the effect of arrival process of node $N_i$ in matrix $A^D$ is restricted to determining the transition probabilities between different $u_i$s, which are specified by matrix $D_i=D_{0,i}+D_{1,i}$, independently. These transitions do not affect the queue length status of other nodes or backoff stages. So, instead of calculating $A^{D} ((\boldsymbol{b}_i,\hat{\boldsymbol{q}}_{-i},u_i),(\boldsymbol{b}'_i,{\hat{\boldsymbol{q}}_{-i}}',u'_i))$ directly, we can solve rate-equivalent network to find $A ((\boldsymbol{b}_i,\hat{\boldsymbol{q}}_{-i}),(\boldsymbol{b}'_i,{\hat{\boldsymbol{q}}_{-i}}'))$ and then multiply it by $D_i(u_i,u'_i)$. This results in $A^D=A \otimes D_i$.
\end{IEEEproof}

Now, by using the above lemma we prove Theorem 5 in the following.
\begin{IEEEproof}[Proof of Theorem 5]
It is clear that $p_{i}^{suc,D}(h)$ which denotes the probability of a successful transmission of node $N_i$ is independent of $u_i$. So, it can be expressed in terms of the probability of a successful transmission in rate-equivalent network, (i.e., $p_{i}^{suc}$) as
\begin{equation}\label{35}
p_{i}^{suc,D}(h)=p_{i}^{suc}(k),
\end{equation}
where $(k-1)c_i+1 \leq h \leq k{c_i}$. It is worth noting that this mapping is with respect to the lexicographic ordering on states, as indicated in Section V. Moreover, it can be concluded from Lemma 1 that
\begin{equation}\label{36}
\sum\limits_{h=(k-1)c_i+1}^{k c_i} {\boldsymbol{\alpha}^{D}(h)}=\boldsymbol{\alpha}(k).
\end{equation}
So,
\ifCLASSOPTIONdraftclsnofoot
\begin{equation}\label{37}
\sum\limits_{h=1}^{m^{D}} {\boldsymbol{\alpha}^{D}(h)p_{i}^{suc,D}(h)}=\sum\limits_{k=1}^{m} {\sum\limits_{h=(k-1)c_i+1}^{k c_i} {\boldsymbol{\alpha}^{D}(h)}p_{i}^{suc}(k)}=
\sum\limits_{k=1}^{m} {\boldsymbol{\alpha}(k)p_{i}^{suc}(k)}=\lambda_{i}^{SR}({\boldsymbol{\lambda}}_{-i}).
\end{equation}
\else
\begin{multline}\label{37}
\sum\limits_{h=1}^{m^{D}} {\boldsymbol{\alpha}^{D}(h)p_{i}^{suc,D}(h)}=\sum\limits_{k=1}^{m} {\sum\limits_{h=(k-1)c_i+1}^{k c_i} {\boldsymbol{\alpha}^{D}(h)}p_{i}^{suc}(k)}=\\
\sum\limits_{k=1}^{m} {\boldsymbol{\alpha}(k)p_{i}^{suc}(k)}=\lambda_{i}^{SR}({\boldsymbol{\lambda}}_{-i}).
\end{multline}
\fi
\end{IEEEproof}
%

%
%

\ifCLASSOPTIONcaptionsoff
  \newpage
\fi



%

\vspace{-0.5cm}
\bibliographystyle{IEEEtran}
\bibliography{references}
\end{document}